\documentclass[final,3p,times]{elsarticle}

\usepackage{amssymb}
\usepackage{amsmath}
\usepackage{lineno}
\usepackage{graphicx} % Required for inserting images

\usepackage[colorlinks=true,breaklinks=true]{hyperref}
\usepackage{changepage} % Para el entorno adjustwidth
\usepackage{tabularx} % Para tablas ajustables
\usepackage{makecell} %para partir lineas en celdas de tablas
\usepackage{pifont} % More styles for bullets
\usepackage{caption}
\usepackage{subcaption}

\usepackage[dvipsnames]{xcolor}

\usepackage[colorlinks=true,breaklinks=true]{hyperref}
\usepackage[normalem]{ulem}
\usepackage[utf8]{inputenc}
\hypersetup{allcolors=[rgb]{0.0 0.0 0.6},linkcolor=[rgb]{0.75 0.05 0.05}}
\usepackage{amsmath,amssymb}
\usepackage{epsfig}  
\usepackage{graphicx}   
\usepackage{slashed}       
\usepackage{tikz}
\usepackage{color}
\usepackage{tikz-feynman}
\usepackage{url}
\usepackage{color}
\usepackage{multirow}
\usepackage{comment}
\usepackage{amssymb}
\usepackage{bm}
\usepackage{footmisc}
\newcommand{\apgcom}[1]{{\textcolor{red}{#1}}}

\newcommand{\beq}{\begin{equation}}
\newcommand{\eeq}{\end{equation}}

\allowdisplaybreaks

\bibpunct{[}{]}{,}{n}{}{,}

\journal{Physics of the Dark Universe}

\begin{document}

\begin{frontmatter}

%\title{ALP-photon conversions in magnetosphere of pulsars \apgcom{perhaps slight change?}}
\title{Gamma Rays from ALP–Photon Conversion and \\ Inverse Compton Reprocessing in Neutron Star Magnetospheres}

\author[inst1,inst2]{F. Giacchino\corref{cor1}}\ead{federica.giacchino@usal.es}\cortext[cor1]{Corresponding author.}

\affiliation[inst1]{organization={Department of Fundamental Physics and IUFFyM, University of Salamanca},addressline={Plaza de la Merced s/n},city={Salamanca},postcode={E-37008},country={Spain}}

\affiliation[inst2]{Istituto Nazionale di Fisica Nucleare - Sezione Roma Tor Vergata, via della Ricerca Scientifica, 00133 Roma, Italy.}

\author[inst1]{C. A. Torres}
\author[inst4,inst5]{G. Galanti}
\author[inst3]{B. J. Kavanagh}
\author[inst1]{M. A. P\'erez-Garc\'ia}
\author[inst3]{J. M. Diego}

\affiliation[inst4]{organization={INAF, Istituto di Astrofisica Spaziale e Fisica Cosmica di Milano},addressline={
Via Alfonso Corti 12}, city={Milano}, postcode={I – 20133}, country={Italy}}

\affiliation[inst5]{organization={INFN, Sezione di Pavia}, addressline={Via Agostino Bassi 6}, city={Pavia}, postcode={I -- 27100}, country={Italy}}

\affiliation[inst3]{organization={Instituto de Fısica de Cantabria (IFCA, UC-CSIC)},addressline={Av. de Los Castros s/n}, city={Santander}, postcode={39005},country={Spain}}

\begin{abstract}
    
Exploring axion-like particle (ALP) signatures from neutron stars (NSs) in the \emph{Fermi}-LAT energy range remains largely unexplored. Neutron stars with exceptionally strong magnetic fields, such as magnetars and pulsars with magnetar-like magnetic fields, provide particularly promising environments for ALP--photon conversion. Magnetars are characterized by surface magnetic fields as large as $B_0\sim(10^{14}$--$10^{15})\,\mathrm{G}$; however, despite their extreme magnetic fields, no steady magnetar emission has been firmly detected in the \emph{Fermi}-LAT energy range, with high-energy activity generally associated with rare flaring episodes.

In this work, we investigate ALP production in the interiors of different classes of NSs and the subsequent conversion of ALPs into photons in their magnetospheres. The ALP emissivity is determined by the stellar density and temperature $T$, while the conversion probability is enhanced by the strong magnetic fields surrounding the star. We further account for photon propagation through the Galactic magnetic field, which can provide an additional contribution to the observable photon flux. We investigate the resulting gamma-ray signatures and assess whether ALP-induced emission from NS magnetospheres could be detectable at energies $E\gtrsim100,\mathrm{MeV}$ in the \emph{Fermi}-LAT band. In addition, we consider if the reprocessing of the magnetospheric photons through inverse Compton scattering can shift part of the emission to higher energies and provide an additional observational signature. We use the resulting fluxes to derive constraints from existing gamma-ray observations and to estimate the sensitivity of future MeV--GeV observations, taking COSI as a representative example.

\end{abstract}
% Keywords
%% Keywords
\begin{keyword}
Axion-like particles (ALP) \sep pulsars \sep neutron stars \sep gamma-ray
%% keywords here, in the form: keyword \sep keyword

%% PACS codes here, in the form: \PACS code \sep code

%% MSC codes here, in the form: \MSC code \sep code
%% or \MSC[2008] code \sep code (2000 is the default)

\end{keyword}

\end{frontmatter}

% T

%%%%%%%%%%%%%%%%%%%%%%%%%%%%%%%%%%%%%%%%%%
% Only
%%%%%%%%%%%%%%%%%%%%%%%%%%%%%%%%%%%%%%%%%%%%%%%%%%
%%%%%%%%%%%%%%%%%%%%%%%%%%%%%%%%%%%%%%%%%%

\section{Introduction}
\label{intro}

Axion-like particles (ALPs) are well-motivated extensions of the Standard Model and represent compelling candidates for new physics, with a rich phenomenology in extreme astrophysical environments~\cite{Arvanitaki:2009fg,Jaeckel_2010,Ringwald:2012hr}. ALPs are pseudoscalar particles inspired by the QCD axion~\cite{Peccei:1977hh, Peccei:1977ur}, originally proposed as a solution to the Strong CP Problem (see e.g.~Ref.~\cite{Hook:2018dlk} for a review). Though they share some properties with axions, ALPs are not constrained to solve the strong CP problem, in which case the mass and couplings of the ALP to fermions and photons can be
freely taken as independent parameters. ALPs belong to the broader class of weakly interacting slim particles (WISPs)~\cite{Arias:2012az}, whose theoretical landscape has recently been systematically catalogued in the WISPedia encyclopedia \citep{Albertus:2026fbe}.
Over the last few years, several complementary laboratory and astrophysical observations have established stringent constraints on the ALP parameter space across multiple coupling channels~\cite{OHare2020}. Regarding the ALP--photon coupling ($g_{a\gamma}$), the CAST experiment, acting as the state-of-the-art axion helioscope, has provided bounds from solar ALPs setting $g_{a\gamma} < 5.7 \times 10^{-11}$~GeV$^{-1}$ for $m_a \leq 0.02$~eV~\cite{CAST:2024eil}. An even tighter constraint in the low-mass regime comes from the non-observation of a gamma-ray burst coincident with Supernova SN1987A, which would be induced by ALP--photon conversion in the Galactic magnetic field. This null result constrains $g_{a\gamma} < 5.3 \times 10^{-12}$~GeV$^{-1}$ for masses $m_a < 4 \times 10^{-10}$~eV~\cite{Payez:2014xsa}. Looking at other sectors, the stellar cooling processes of red giant branch stars, such as $\alpha$-Centauri~\cite{Capozzi:2020cbu}, impose the most restrictive upper bound on the axion-electron coupling, yielding $g_{ae} < 1.5 \times 10^{-13}$~GeV$^{-1}$. Concurrently, the neutrino emission duration from SN1987A restricts the anomalous cooling mediated by axion emission via nucleon couplings, establishing that $g_{ap} \sim g_{an} < 10^{-9}$~GeV$^{-1}$ \cite{Lella:2023bfb, Carenza:2019pxu}.

In a more extreme stellar scenario, neutron stars (NSs) provide unique conditions of high density and extreme magnetic field strength, making them ideal laboratories to investigate ALP production and conversion mechanisms (see e.g.~Refs.~\cite{Pshirkov:2007st,Safdi:2018oeu,Hook:2018iia,Huang:2018lxq,Lloyd:2020vzs,Edwards:2020afl,Agrawal:2021dbo,Fortin:2021sst,Battye:2021xvt,Noordhuis:2023wid,Roy:2025mqw,Bhura:2026bpy}). Magnetars represent a subclass of neutron stars characterized by ultra-strong surface magnetic fields, $B_0 \sim 10^{14}$--$10^{15}$~G, which exceed those of ordinary radio pulsars by several orders of magnitude \cite{2004hpa..book.....L}. % While their extreme magnetic environments make them highly appealing for studying the $B^2$-dependent ALP--photon mixing, searches for persistent high-energy gamma-ray emission from canonical magnetars with the \emph{Fermi} Large Area Telescope (\emph{Fermi}-LAT) have so far yielded only null results.  
Their extreme magnetic environments make them promising targets for searches of ALP-induced gamma-ray emission, since ALP-photon conversion in strong magnetic fields can produce detectable high-energy photons. While these objects are therefore appealing laboratories to study the $B^2$-dependent ALP-photon mixing, searches for persistent high-energy gamma-ray emission from canonical magnetars with the \emph{Fermi}-LAT have so far yielded only null results. %\bjk{<- We haven't said yet that we're interested in looking for ALP-induced gamma-ray emission! I would make this clear (i.e. that we're interested in gamma rays) before we get into the limitations/fluxes from magnetars.} 
Dedicated analyses by the \emph{Fermi}-LAT Collaboration in 2010 and 2017 found no evidence of steady, persistent gamma-ray flux above 100~MeV from these sources \cite{Fermi-LAT:2010Magnetars, Li:2017Magnetars}. Their baseline emission is dominated by soft X-rays and soft gamma rays powered by magnetic dissipation rather than rotational losses. Although transient gamma ray emission up to 10~GeV has been occasionally reported during anomalous flaring episodes---such as a recent $4.4\sigma$ excess from 1E~1048.1--5937 following an X-ray burst \cite{Ramakrishnan:2024pmg}---persistent emission remains undetected, heavily obscured or suppressed by efficient non-linear magnetospheric mechanisms like photon splitting and pair production ($e^\pm$). Consequently, canonical quiescent magnetars do not provide a viable, clean window for testing steady ALP-induced gamma ray fluxes within the \emph{Fermi}-LAT energy range ($\sim 100$~MeV to $1$~TeV).

The recent observational and theoretical evidence suggests that magnetars and standard pulsars do not constitute two disconnected classes but rather form a physical continuum within the broader neutron star population \cite{Kaspi:2017fwg, Esposito:2018gvp}. The transition between these regimes depends on the interplay between magnetically driven and rotation-powered energy losses. 

Driven by this paradigm, this work focuses on a systematic comparison of the expected gamma-ray fluxes from ALP production and subsequent ALP-photon conversion across different categories of rotation-powered objects, specifically selecting young, highly energetic radio pulsars and \emph{magnetar-like pulsars}. The latter represent hybrid transitional objects, such as PSR~J1119--6127, which exhibit high surface magnetic fields ($B \geq 4 \times 10^{13}$~G) and sporadic magnetar-like outbursts, while maintaining a stable, rotation-powered gamma-ray profile natively detectable by \emph{Fermi}-LAT. This comparative approach allows us to investigate how different combinations of spin period ($P$), magnetic field strength ($B_0$), and age govern the resulting ALP signatures.

Following the state-of-the-art data from the Fourth \emph{Fermi}-LAT Source Catalog (4FGL-DR4) \cite{Ballet:2023qzs} and the Third Pulsar Catalog (3PC), \cite{Fermi-LAT:2023zzt} we apply this framework to four target pulsars representing distinct astrophysical categories, whose main timing, spatial, and magnetospheric parameters are summarized in Table \eqref{tab:pulsars_transposed_ul}. Specifically, our selection spans three representative classes of objects designed to probe different energetic regimes: a young canonical rotation-powered gamma-ray pulsar (\emph{the Crab}, PSR~J0534+2200); a transitional magnetar-like pulsar embedded in a prominent pulsar wind nebula (PSR~J1846-0258); a clear, high-field magnetar-like pulsar exhibiting sporadic magnetospheric outbursts (PSR~J1119-6127); and a highly distant, older pulsar configuration (PSR~J1341-6220).

\begin{table}
\centering
\small
\caption{Main parameters from 4FGL-DR4 \cite{Ballet:2023qzs} and 3PC \cite{Fermi-LAT:2023zzt} of the three pulsars detected by \textit{Fermi}-LAT. Age [years], Energy Flux [erg/cm$^2$/s] Pivot energy [MeV], Distance [kpc], pulsation period [ms], surface magnetic field $B_0= 3.2 \times 10^{19} \sqrt{P\dot{P}}$ [G], surface electron number density  from the GJ model $n_{e,0}\simeq \boldsymbol{\Omega} \cdot \mathbf{B} / (2\pi e c)$ [cm$^{-3}$], $95\%$ confidence level semi-major axis of the LAT error ellipse [arcmins]. The last object is a composite PWN+pulsar source. }

\setlength{\tabcolsep}{4pt}

\begin{tabular}{l c c c c}
\hline
\hline
Parameter & J0534+2200 (Crab) & J1119-6127 & J1341-6220 & J1846-0258\\
\hline
4FGL NAME & 4FGL J0534.5+2200 & 4FGL J1119.1-6127 & 4FGL 1341.7-6216 & 4FGL J1846.9-0247c\\
RA & 83.63308 & 169.80958 & 205.42763 & 281.60208\\
Dec & +22.01447 & -61.46375 & -62.33908 & -2.97444 \\
TS & 48053.3 & 1292.9 & 153.4 & 100 \\
Age & 1268.79 & 1603.49 & 12112.6 & 727.98026 \\
Spectrum Type & PLEC & PLEC & PLEC & PL (PWN-dominated)\\
Energy Flux ($10^{-11}$) & $150\pm 1.3$ & $4.4\pm 0.32$ & $2.2\pm 0.52$ & $4.5\pm 1.5$ \\
Pivot Energy & 1592.41 & 2053.90 & 3156.22 & unknown\\
Distance & $2.0^{+0.4}_{-0.3}$ & $8.4\pm0.4$ & $12.6\pm 5.0$ & $5.8$ \\
Variability Index & $141.34$ & $16.90$ & $10.12$ & unknown \\
Pulsation period & $33.7$ & $409.1$ & $193.4$ & $326.571$\\
$dP/dt$ & $4.2\times10^{-13}$ & $4.04 \times 10^{-12}$ & $2.53\times 10^{-13}$ & $7.11\times 10^{-12}$ \\
Conf. 95$\%$ & $0.0073$ & $0.0235$ & $0.0612$ & unknown \\
\hline
$B_0$ & $3.81\times 10^{12}$ & $4.11\times 10^{13}$& $7.08\times10^{12}$ &$4.88\times 10^{13}$\\
$n_{e,0}$ & $7.85\times 10^{16}$ & $7.00\times 10^{16}$ & $2.54\times 10^{16}$ & $1.04\times 10^{17}$\\
\hline
\hline
\end{tabular}

\label{tab:pulsars_transposed_ul}
\end{table}

We evaluate the equatorial surface magnetic field assuming a standard rotating vacuum dipole configuration, given by $B_{\mathrm{s}} \equiv B_0 = 3.2 \times 10^{19} \sqrt{P\dot{P}}$~G where \(P\) is the spin period and \(\dot{P}\) its time derivative. Close to the surface, the charge distribution is approximately governed by the Goldreich-Julian (GJ) model \cite{GJ1969ApJ...157..869G}, providing a characteristic corotating primary plasma number density $n_{e,0} \simeq \boldsymbol{\Omega} \cdot \mathbf{B} / (2\pi e c)$ where \(e\) is the elementary charge and \(c\) is the speed of light.

Within our theoretical framework, ALPs ($a$) are efficiently produced in the NS core and crust via multiple channels through independent couplings as we will describe later on. These include nucleon ($N$)-induced processes in the dense core, such as nucleon--nucleon bremsstrahlung ($N + N \rightarrow N + N + a$) and pion-induced thermal reactions, alongside lepton-induced channels in the stellar crust, dominated by electron--ion bremsstrahlung ($e + Z \rightarrow e + Z + a$), and thermal Primakoff processes, see for example \cite{Raffelt_2024,Li_2024}.

The microscopic ALP emission rates ($Q_a$) for core processes exhibit a highly sensitive power-law dependence on the local core temperature, scaling typically as $Q_a \propto T_{\rm core}^6$. Consequently, the core temperature represents the primary driver of the initial axion flux. In standard neutron star cooling theory where neutrino emission dominates the early $\sim10^6$ yr thermal evolution, the core temperature can be approximated as a function of the characteristic NS age $t_{\rm NS}$ via the standard relation \cite{yakovlev2004neutron}
\begin{equation}
T_{\rm core} \sim 10^9 \text{ K} \times \left( \frac{10^3 \text{ yr}}{t_{\rm NS}} \right)^{1/6}=0.086 \text{ MeV}\times \left( \frac{10^3 \text{ yr}}{t_{\rm NS}} \right)^{1/6}\,.
\label{temperature}
\end{equation}

Applying this analytical model to our selected targets in Table \eqref{tab:pulsars_transposed_ul}, we derive the following core temperatures: PSR J0534+2200 (Crab): $T_{\rm core} \sim 1.0 \times 10^9$~K ($0.086$ MeV), PSR J1119-6127: $T_{\rm core} \sim 9.4 \times 10^8$~K ($0.081$ MeV), PSR J1341-6220: $T_{\rm core} \sim 5.4 \times 10^8$~K ($0.0465$ MeV), PSR J1846-0258: $T_{\rm core} \sim 1.0 \times 10^9$~K ($0.086$ MeV). This comparison reveals that young pulsars share nearly identical internal core temperatures ($\sim 10^9$~K) with typical young magnetars, despite exhibiting vastly different surface X-ray properties ($T_{\rm surf}$). While magnetars possess enhanced surface temperatures ($T_{\rm surf} \sim 7 \times 10^6$~K) due to magnetic heating mechanisms in the outer crust, the core thermal state remains uniform across the youngest population. Because ALP production takes place in the deep volume, it is entirely dictated by $T_{\rm core}$, ensuring comparable production rates between young high-field pulsars and magnetars. 

Once produced, these relativistic ALPs with feeble interactions with ordinary matter escape the dense stellar medium and undergo mixing with photons and subsequent Inverse Compton processes. We account for the total conversion probability by tracking the ALP stream through two distinct regions: first, locally within the external magnetosphere where the intense dipole field $B(r) \propto r^{-3}$ triggers localized conversion, and subsequently over kiloparsec scales through the regular and turbulent components of the Galactic magnetic field \cite{Jansson_2012}, which acts as a macroscopic coherent amplifier for low-mass ALPs ($\leq10^{-3}$ eV).

Since the photons produced through ALP conversion and subsequent Inverse Compton scattering are expected to populate the hard X-ray and MeV gamma-ray bands, as the ALPs are produced with characteristic energies comparable to the neutron-star core temperature and ALP-photon conversion approximately preserves their energy, instruments operating in this energy range are particularly well suited to test the scenario. %\bjk{Why are they expected to populate these bands? I guess we should make the logic clear (light ALPs are produced with an energy comparable to the core temperature, which when converted into photons corresponds to energies in the X-ray and gamma-ray bands).} 
We therefore consider archival observations from COMPTEL \cite{Schoenfelder:1993COMPTEL}, %and INTEGRAL~\cite{Winkler:2003INTEGRAL}
which provide existing constraints on persistent emission from neutron stars, and evaluate the prospects for the Compton Spectrometer and Imager (COSI) \cite{Kierans:2022eid,Beechert:2022phz,COSI:2026asd}, whose substantially improved sensitivity in the MeV domain is expected to probe previously unexplored regions of the ALP parameter space.  Other missions, such as the proposed newASTROGAM are also well suited for this task \cite{Berge:2025kff}.

%To benchmark our findings against actual high-energy gamma-ray data, Fig.~\ref{fig:sed_PSR1119} displays the Spectral Energy Distribution (SED) of PSR J1119-6127 extracted from the \emph{Fermi}-LAT data, providing the empirical flux baseline and the corresponding upper limits ($TS < 9$) used to bound the ALP parameter space.

%\begin{figure}[h!] \centering \includegraphics[width=1.0\linewidth]{sed_plot.png}\caption{Spectral Energy Distribution (SED) of the magnetar-like pulsar PSR J1119-6127 with 60 bins per decade. Upper limits (UL) correspond to bins with a $TS < 9$, establishing the background baseline in the \emph{Fermi}-LAT energy window.} \label{fig:sed_PSR1119}
%\end{figure}

Regarding ALP decay, for photons with energies \(E_\gamma \gtrsim 100\,\mathrm{MeV}\), the parent ALP would typically need a mass \(m_a \simeq 2E_\gamma \gtrsim 200\,\mathrm{MeV}\) implying a comparatively heavy ALP. Since the decay width scales as  \(\Gamma_{a\gamma\gamma}\propto g_{a\gamma}^2 m_a^3\)~\cite{Jaeckel_2010}, such heavy ALPs decay very rapidly so that they are not relevant for the scenarios investigated here.

This work is structured as follows. In Sec.~\eqref{section:ALPproduction} we compute the main ALP production processes in the NS involving nucleon and lepton populations. Later, in Sec.~\eqref{section:lpconversion} we treat the ALP-photon conversion inside and outside the magnetosphere showing the fluxes for the four target PSRs in Table \eqref{tab:pulsars_transposed_ul}. In Sec.~\eqref{section:IC} we further consider the energetic shift from the Inverse Compton (IC) process for the photons converted inside the magnetosphere including the loss of efficiency factors from pre-IC absorption. Finally in Sec.~\eqref{section:results} we calculate the expected flux and show the constraints in the ALP parameter space arising from archival COMPTEL data and projections for future MeV experiments such as COSI. %, and only for the Crab we test the parameter space with the 
We discuss further observational prospects to finally conclude in Sec.~\eqref{section:conclusion}.

%%%%%%%%%%%%%%%%%%%%%%%%%%%%%%%%%%%%%%%%%%
\section{ALP effective Lagrangian and production channels}
\label{section:ALPproduction}

Rather than specifying a particular ultraviolet completion, we work within a low-energy effective field theory (EFT) describing the interactions of a light pseudoscalar ALP, $a$, with Standard Model fermions, $f$, and photons. Thus we consider the interactions of ALPs with leptons, nucleons and photons in dense, charge-neutral stellar matter.  The relevant effective Lagrangian can be written as~\cite{Georgi:1986df}.
\begin{equation}
\mathcal{L} \supset 
\frac{1}{2} (\partial_{\mu} a)(\partial^{\mu} a)
-\frac{1}{2}m_a^2 a^2
+\frac{1}{4}g_{a\gamma}a F_{\mu\nu}\tilde{F}^{\mu\nu}
+\sum_{f=e,n,p}\frac{g_{af}}{2}(\partial_{\mu} a)\bar{f}\gamma^{\mu}\gamma^5 f \,,
\label{eq:lagrangian}
\end{equation}
where $\tilde{F}^{\mu\nu}=\frac{1}{2}\epsilon^{\mu\nu\alpha\beta}F_{\alpha\beta}$ is the dual Faraday tensor, with $\epsilon^{0123}=-1$ the completely antisymmetric Levi-Civita tensor. The ALP mass $m_a$ and the couplings $g_{a\gamma}$ and $g_{af}$, to photons and fermions, respectively, are treated as independent parameters. These interactions arise from dimension-5 operators and therefore the couplings carry dimensions of inverse mass in $c=1$ units (GeV$^{-1}$). The derivative fermion coupling in Eq.~\eqref{eq:lagrangian} can be rewritten as $-i g_{af} m_f a \bar f \gamma_5 f$ after integrating by parts and using the fermion equations of motion (see~\cite{Darme:2020sjf}). Photon-ALP interactions gives rise to two effects \cite{Raffelt:1987im,Raffelt:1996wa}: photon-ALP oscillations, and change of the photon-polarization state, which are potentially detectable in astrophysical spectra~\cite{DeAngelis:2007dqd,Hochmuth:2007hk,Sanchez-Conde:2009exi} and on photon polarization~\cite{Galanti:2024lfn}.

In dense stellar environments, several processes involving nucleons, leptons, and photons can contribute to the production of ALPs.  The total differential axion emissivity\footnote{The emissivity is 
the local energy-loss rate per unit volume due to axion emission, $Q \equiv \frac{dE}{dV\,dt}$.
% \bjk{I'm a bit confused. Is this the same emissivity as in Eq. (3) above? Why is there a different symbol? In any case, it would be good to define the emissivity further up, around Eq. (3).}
} $\frac{dQ_a}{dx}$ can be written as the sum of the contributions from the relevant production channels,
\begin{equation}
\frac{dQ_a}{dx}=\sum_{i}\frac{dQ_{a,i}}{dx},
\end{equation}
where $x=E_a/T$ denotes the dimensionless ALP energy to temperature ratio (we take $k_B=1$). In what follows we describe % We label $i=nn,pp,np,\pi,ep,eZ,\gamma$ for 
the different $i-$indexed processes we consider involving nucleons ($N\equiv n,p$), leptons ($e$), photons $\gamma$, pions and nuclei with charge $eZ$. 
Typically,  as done in \cite{Fortin:2021sst}, we write the differential emissivity factorized into the integrated emissivity and a dimensionless spectral function $\frac{dQ_{a,i}}{dE_a}=\frac{Q_{a,i}}{T}\Phi_i(x)$. $\Phi(x)$ describes the spectral shape and is normalized such that $\int_0^{\infty} d x\, x \,\Phi(x)=1$.

%\begin{equation}\frac{dQ}{dx}|_{NN}=\sum_{i}\frac{dQ_i}{dx}= Q_i\times N_i\frac{x^3(x^2+4\pi^2)}{e^x-1},\quad \textit{with}\,\,x=E_n/T\label{diffemissivity}\end{equation} with $i=nn,pp,np$, $N_{nn}=N_{pp}=315/(124\pi^6)$ and $N_{np}=315/(31\pi^6)$, and $E_n$ the energy of the emitted axion. \textbf{HAS TO PUT ALSO THE PION-AXION CONVERSION}. 

The differential ALP number spectrum produced inside the neutron-star interior is obtained by integrating the local emissivity over the stellar volume,
\begin{equation}
\frac{dN_a}{dE}
=
\frac{1}{E}\frac{dL_a}{dE}
=
4\pi\int_0^R dr\, r^2\,\frac{dQ_a(r)}{dx}\,\frac{1}{x\,T(r)^2},
\end{equation}
where \(R\) is the stellar radius\footnote{In a general-relativistic treatment, using a static spherically symmetric metric $ds^2=-e^{2\Phi(r)}dt^2+e^{2\Lambda(r)}dr^2+r^2 d\Omega^2$, the Newtonian volume element \(4\pi r^2dr\) should be replaced by the proper-volume element \(4\pi r^2 e^{\Lambda(r)}dr\), and the ALP energy must be redshifted according to \(E_\infty=e^{\Phi(r)}E\).
In what follows we will omit these corrections as they will only modify to factor $\mathcal{O}(1)$.}. Generally speaking, the differential emissivity includes the appropriate kinematic phase space factors, as well as the quantum statistical occupation and blocking factors associated with the microscopic production processes considered. It also incorporates the relevant neutron star properties, such as the local density profile. In particular, %\bjk{Maybe it's worth making it very explicit here that the (differential) emissivities $Q$ depend on the NS density. I was initially expecting some information about the fermion density to appear in the integral in Eq. (4), but I now realise that this is contained in the definition of the emissivity, so maybe we can add a phrase to make this clear. Then it also becomes clear why we're talking about the density profile in the next paragraph.}
finding a fully consistent density profile for a neutron star with given mass $M$ and radius $R$, compatible with a nucleonic equation of state (EoS) across all regions, is a non-trivial task, especially in the high-density core \cite{Fortin2016PhRvC..94c5804F}. The density and composition profiles adopted here are obtained from solving the structure equations under the PAL EoS \cite{Prakash1988PAL} and are intended to illustrate the typical radial behavior obtained from modern unified nucleonic equations of state, rather than to represent a specific microscopic EoS.
Charge neutrality is fulfilled by construction, so that in a simplified $n-p-e$ system number densities are such  $n_e = n_p$ at all stellar radii. We summarize the typical order of magnitude of particle number densities for a nucleonic, beta-equilibrated neutron star with $M = 1.4\,M_\odot$ and $R = 13\,\mathrm{km}$. In the inner core, electron and proton densities are $n_e \sim n_p \sim \mathcal{O}(10^{37})\,\mathrm{cm^{-3}}$, while neutrons reach $n_n \sim \mathcal{O}(10^{38})\,\mathrm{cm^{-3}}$. Towards intermediate radii, densities decrease by one to two orders of magnitude, with $n_e \sim n_p \sim \mathcal{O}(10^{35\text{--}36})\,\mathrm{cm^{-3}}$ and $n_n \sim \mathcal{O}(10^{37})\,\mathrm{cm^{-3}}$. Near the stellar surface, they drop further to $n_e \sim n_p \sim \mathcal{O}(10^{34})\,\mathrm{cm^{-3}}$ and $n_n \sim \mathcal{O}(10^{36})\,\mathrm{cm^{-3}}$, eventually becoming negligible at the outer boundary, see \cite{Haensel:2007yy}.

%We stress that the tabulated density profile is valid only for the specific mass and radius adopted in this work, as neutron stars with different global properties may exhibit substantially different internal structures.

In the following subsections, we outline the production processes we consider in our model when exploiting nucleon-, electron-, and photon-ALP couplings.

\subsection{ALP nucleon coupling: nucleon bremsstrahlung}

We consider nucleon bremsstrahlung taking place by the process
\begin{equation}
    N+N'\rightarrow N+N'+a\,,
\end{equation}
with $N,N'=n,\,p$. Inside the NS core nucleon matter is so dense and degenerate that this has been quoted as the main production mechanism of ALPs~\cite{Raffelt:1996wa}. The dominant contribution to the bremsstrahlung rate comes 
from nucleons near their Fermi surface to prevent full final phase space blocking\footnote{See for example, computations in  \cite{Carenza:2019pxu} where it is assumed that the nucleons participating in the process are located near the Fermi surface.}
%, considering the matter inside as effectively at $T \simeq 0$, while the nucleons at the surface have $T \neq 0$.}, 
since the core temperature is much smaller than the nucleon Fermi energy. This reflects Pauli blocking in the degenerate core: only nucleons within $\sim k_B T$ of the Fermi surface have access to unoccupied final states. Following \cite{Fortin:2021sst} the emissivity for $n + n \rightarrow n + n + a$ is given by
\begin{equation}
Q_{a,nn} = \frac{31}{2835\pi} C_\pi 
\left( \frac{m_n}{m_\pi} \right)^4 f^4\, g_{an}^2\, 
p_{F_n} F(d_n) \,T^6 ,
\label{eq:emissivity_nn}
\end{equation}
where where \(f\equiv f_{\pi NN}\simeq 1\) denotes the dimensionless
pseudovector pion--nucleon coupling constant $d_n = \frac{m_\pi}{2 p_{F_n}}$, $p_{F_n}$ is the neutron Fermi momentum, and the 
function $F(d_n)$ is defined as
\begin{equation}
F(d_n) = 4 - \frac{1}{1 + d_n^2} - 5d_n \arctan\left( \frac{1}{d_n} \right)
+ 2d_n^2 \sqrt{1 + 2d_n^2}\, \arctan\left( \frac{1}{\sqrt{1 + 2d_n^2}} \right).
\label{eq:Ffunction}
\end{equation}
Here, $C_\pi$ represents the correction factor that accounts for the overestimate of the 
strong interaction rate in the one-pion exchange approximation (OPE). In the OPE limit, 
$C_\pi = 1$, while a more realistic estimate gives $C_\pi = 1/4$ as discussed in 
Appendix~C of \cite{Fortin:2021sst} and also described in \cite{Carenza:2024ehj} and \cite{Carenza:2020cis}. The neutron Fermi momentum is given by $p_{F_n} = \hbar (3 \pi^2 n_n(r) )^{1/3}$, where $n_n$ is the neutron number density. Note that in \cite{Lloyd:2020vzs,Fortin:2021sst},  the matter density is assumed to be uniform within the core while here,
%up to $R_{\rm core}=11.3\,\mathrm{km}$. 
instead, we adopt a number density profile that varies with the NS radius.

For the process $p + p \rightarrow p + p + a$, the emissivity expression is analogous to Eq.~\eqref{eq:emissivity_nn}, with $p_{F_n}$ replaced by the proton Fermi momentum $p_{F_p}$ and 
$d_n$ by $d_p = m_\pi / (2 p_{F_p})$, namely

\begin{equation}
Q_{a,pp} = \frac{31}{2835\pi} C_\pi 
\left( \frac{m_n}{m_\pi} \right)^4 f^4 g_{ap}^2 
p_{F_p}(r) F(d_p) T^6 ,
\label{eq:emissivity_pp}
\end{equation}
where $p_{F_p} = \hbar (3 \pi^2 n_p(r))^{1/3}$ is the proton Fermi momentum.
We write
$\Phi(x)=\frac{x^4 e^{-x}}{48}$ for $Q_{a,pp}$, $Q_{a,nn}$ and $Q_{a,np}$.

Finally, for the mixed process $n + p \rightarrow n + p + a$ (assuming in the NS neutron rich environment $p_{Fn}>p_{Fp}$), one obtains \cite{Fortin:2021sst}
\begin{equation}
Q_{a,np} = \frac{124}{2835\pi} C_\pi 
\left( \frac{m_n}{m_\pi} \right)^4 f^4 g_{an}^2 
p_{F_p}(r)  \,G(d_n,d_p) T^6 ,
\label{eq:emissivity_np}
\end{equation}
where 
\begin{equation}
\begin{aligned}
G(d_n,d_p)=&\,5
+\frac{1}{2}\frac{d_p^{2}}{1+d_p^{2}}
+d_p^{4}\left(\pi-\arctan\!\left(\frac{2d_p}{1-d_p^{2}}\right)\right)
-4d_p\arctan\!\left(
\frac{4d_n^{2}d_p}
{4d_n^{2}d_p^{2}+d_p^{2}-d_n^{2}}
\right)\\[2mm]
&-\frac{(d_n-d_p)^{3}}
{d_n\left(4d_n^{2}d_p^{2}+(d_n-d_p)^{2}\right)}
-\frac{(d_n+d_p)^{3}}
{d_n\left(4d_n^{2}d_p^{2}+(d_n+d_p)^{2}\right)}\\[2mm]
&-\frac{4d_n^{2}d_p^{3}}{H(d_n,d_p)}
\left[
\pi+
\arctan\!\left(
\frac{
2d_p\left(d_p^{2}(1+4d_n^{2})-d_n^{2}\right)H(d_n,d_p)
}{
(1+4d_n^{2})^{2}d_p^{6}
-(1+10d_n^{2}+40d_n^{4})d_p^{4}
+d_n^{2}(2-7d_n^{2})d_p^{2}
-d_n^{4}
}
\right)
\right]\,,
\end{aligned}
\end{equation}

and 

\begin{equation}
H(d_n,d_p) = 
\sqrt{(1 + 8d_n^{2} + 32d_n^{4})\,d_p^{4} 
+ 2d_n^{2}(4d_n^{2} - 1)\,d_p^{2} 
+ d_n^{4}}\, .
\end{equation}
%\bjk{Is it worth adding the explicit expressions for these $F$, $G$, $H$ functions in an appendix? They break up the flow of the text here, and they're a little hard to read. Equally, there are no so many of these functions, so maybe they're okay here (and maybe having a whole appendix just for that is too much...)}

%For this case the previous rule with differential emissivity holds.
%\apgcom{I think the differential values are clear...but please, ALL check.} \fcom{CHECKED}\fcom{In \cite{Bottaro:2024ugp}
%
In the NS core, neutron and proton superfluid pairing is expected to occur in the $^1S_0$ and $^3P_2$ channels (for neutrons) and $^1S_0$ channel (for protons), see \cite{Dean:2002zx}. For simplicity we have not included the supefluid corrections of hadron species as in \cite{Fortin:2021sst}. As we do not keep track of dynamical evolution in our model, we assume the core has already thermalized, i.e.\  with uniform temperature, for currently known magnetars \cite{Pons:2019zyc}.
%In our analysis, we consider \textbf{the Fermi momenta $p_{F_n} \simeq 340~\mathrm{MeV}$ and $p_{F_p} \simeq 60~\mathrm{MeV}$},corresponding to beta-equilibrated, charge-neutral nuclear matter in the magnetar core, as obtained for densities of a few times the nuclear saturation density ($n_B \simeq 3n_0$) using realistic equations of state such as the IU-FSU (IUF) model, and consistent with previous works by Iwamoto~\cite{Iwamoto:1984ir}, Raffelt and Seckel~\cite{Raffelt:1995}, and Yakovlev et~al.~\cite{Yakovlev:2001}.

\subsection{ ALP nucleon coupling: pion-axion conversion}

Some recent studies \cite{PhysRevC.101.035809, Harris:2024ssp} demonstrated that the strong pion-nucleon interactions enhance the abundance of negatively charged pions most notably at temperatures of a few MeV. Thus the pion population grows with temperature and as density increases beyond saturation density, surpassing the lepton population.  In \cite{Carenza:2020cis, Carenza:2024ehj} the ALP emissivity from the following process was considered
%(\textbf{In this paper they consider as astrophysical environment core-collpse SN or neutron star merger - I can use it also for magnetars or pulsar magnetar-like?Here in \cite{Lella:2022uwi} they talk about the protoneutron star}) 
\begin{equation}
    p+\pi^-\rightarrow n+a,
    \label{pionprocess}
\end{equation}
involving thermal pions. For the reaction in Eq.~\eqref{pionprocess}, chemical equilibrium requires $\mu_{\pi}=(\mu_n+\mu_a-\mu_p)$, with the very light ALP satisfying $\mu_a\ll\mu_N$. Then from \cite{PhysRevC.101.035809} we can see that strong pion-nucleon interactions can enhance the thermal population of negatively charged pions at sufficiently high densities and temperatures. The pion fugacity, defined as
\[
z_\pi = \exp\!\left(\frac{\mu_\pi - m_\pi}{T}\right),
\] measures the proximity of the pion gas to chemical degeneracy. If $\mu_\pi - m_\pi > 0$, at a given density inside the star, then a pion population can coexist with baryons. %Typical values quoted in the literature are $Y_\pi\sim 0.1$, with $Y_\pi=n_{\pi^{-}} / n_B$, where $n_{\pi^{-}}$ is the pion number density and $n_B$ is the baryon number density. \bjk{It's not clear why we're introducing this $Y$ parameter. It doesn't seem to relate to the discussion so far.}
%the exponent is negative and $z_\pi \ll 1$ (as it is showed in Fig.\ref{fig:fugacity}), implying that the pion population is exponentially suppressed. Physically, this corresponds to a dilute Boltzmann regime, where pions are far from condensation and their number density is thermally suppressed.

The expression for the emissivity of the process in Eq.~\eqref{pionprocess} is given by \cite{Carenza:2024ehj} 
\begin{equation}
Q_{a,\pi} =
\frac{30}{\pi^{2}} \,
\left[
\frac{1}{2}\left(g_{an}^{2} + g_{ap}^{2}\right)
+ \frac{1}{3} g_{an} g_{ap}
\right]
\left(\frac{g_{A}}{f_{\pi}}\right)^{2}
n_{p}(r)\, z_{\pi}\, T^6,
\end{equation} 
where $g_{an}$ and $g_{ap}$ denote the axion couplings to neutrons and protons, respectively; $g_{A}=1.27$ is the axial-vector coupling constant; and $f_{\pi}=92.4$ MeV is the pion decay constant. In this case 
\begin{equation}
\Phi_\pi(x)=\frac{x^4}{24 \zeta(5)} \frac{1}{e^x-1}\,,
\end{equation}
with $\zeta(5)\simeq$ 1.04.
%and we now write
%\begin{equation}
%\frac{d Q_{a,\pi}}{d x}=Q_\pi \frac{x^4}{24 \zeta(5)} \frac{1}{e^x-1} \, 
%\label{depion}
%\end{equation}
%\bjk{I guess once we've explained (further up) that we write the emissivity as an integrated emissivity and a dimensionless spectral function, we don't need to write this equation explicity.}
% \bjk{I commented this before, but I think it still stands - why are we writing this in full again? Once we specify $Q$ and $\Phi(x)$, we don't need to write out $dQ/dx$ in full. And I think it's worth cutting down the number of equations (especially if they're not needed).}

\subsection{ ALP electron coupling:  electron–proton bremsstrahlung}

This process reads
\begin{equation}
e^{-}+p \rightarrow e^{-}+p+a,
\label{ep}
\end{equation}
%If electrons are degenerate (or muons) then similar expressions apply (see \cite{Iwamoto:1984ir, Raffelt:1996wa}):
%
%\textbf{I did not find this formula in the references. It ha dimension $[E]^3$, we expect dimension $[E]^5$. This process occurs only in the crust or in all NS?}
%
and contributes mainly in the stellar core where protons are not bound in nuclei. Following \cite{Iwamoto:1984ir}, for ultra-relativistic and degenerate electrons, the emissivity is
\begin{equation}
Q_{a,ep} = \frac{\pi^2}{120} \alpha^2 g_{ae}^2 \frac{m_e^2}{p_{F,e}^2} n_p T^4 L_{ep},
\end{equation}
where $n_p$ is the proton number density and $p_{F, e}=\left(3 \pi^2 n_e(r)\right)^{1 / 3}$ is the electron Fermi momentum. The Coulomb logarithm for relativistic electrons is given by
\begin{equation}
L_{ep} = \ln \left( \frac{2 p_{F,e}}{m_D} \right) - \frac{1}{2},
\end{equation}
and the electron screening is described by the relativistic Thomas--Fermi (Debye) mass $m_D$ given by   
$m_D^2 = \frac{4\alpha}{\pi} p_{F,e}^2$~\cite{Iwamoto:1984ir}.
The Coulomb logarithm term is therefore approximated as $L_{ep} \approx \frac{1}{2}\ln(\pi/\alpha) - \frac{1}{2} \approx 2.5$. The emissivity scales as $T^4$, consistent with the phase space available for degenerate electrons scattering off partially degenerate protons. 

%Thus the differential emissivity is
%\begin{equation}
%\frac{d Q_{a,e p}}{d x}=Q_{a,e p} \Phi(x)\,,
%\end{equation}
%using $\Phi(x)=\frac{15}{\pi^4}\frac{x^3}{e^x-1}.$ \bjk{Same comment as in the previous section - I would remove the explicit equation for $\frac{d Q_{a,e p}}{d x}$, because we already have $Q$ and $\Phi$ given explicitly. }

\subsection{ ALP electron coupling: electron–ion bremsstrahlung}

In this process electrons scatter off ions with charge $eZ$ mostly in a warm fluid or crystallized system
\begin{equation}
e^{-}+Ze \rightarrow e^{-}+Ze+a\,,
\end{equation}
with  electrons remaining degenerate and relativistic. The emissivity is \cite{Iwamoto:1984ir}
%
%\begin{equation}
%Q_{e Z}=\frac{\alpha^2 g_{a e}^2}{120 \pi^3} \frac{Z^2 n_Z}{m_Z^2} p_{F, e} T^6 \Lambda_{e Z}
%\end{equation}
%
\begin{equation}
    Q_{a,eZ}=\frac{\pi^2}{120}Z^2\alpha^2\frac{g_{ae}^2m_e^2}{p^2_{F,e}}n_Z T^4\Lambda_{eZ},
\end{equation}
with $n_Z$ the density of single ion species of charge $eZ$ and the Coulomb logarithm for the periodically arranged crystal is 
\begin{equation}
\Lambda_{e Z}=\int_0^{2 p_{F,e}} \frac{d q}{q} \frac{S(q)|F(q)|^2}{\left(1+q^2 / m_D^2\right)^2}\,.
\end{equation}
The evaluation of the Coulomb logarithm $\Lambda_{e Z}$ accounts for the static structure factor $S(q)$, which describes spatial correlations among ions in the NS crust as shown in the pioneering work by Horowitz et al.~\cite{horoPhysRevC.69.045804}, and the finite-size effects via the nuclear form factor $F(q)$ as shown in, for example, \cite{Haensel2007, Pons2019}. Assuming a uniformly charged spherical nucleus, the form factor is given by
\begin{equation}
F(q)=3\,\frac{\sin(qR_N)-qR_N\cos(qR_N)}{(qR_N)^3},
\end{equation}
where the effective nuclear radius is parametrized as
$R_N = 1.2\,A^{1/3}\ \mathrm{fm}
$, with $A$ the nuclear mass number \cite{Pons2019}.
This approximation provides an adequate description of nuclear charge
distributions for the purposes of electron--ion scattering in the crust. Charge neutrality implies that the electron number density is related to the
ion density by
$n_e = Z\,n_i$.
%, where $Z$ is the nuclear charge.
%Electrons are assumed to be strongly degenerate and ultrarelativistic, so that
%their Fermi momentum given by the usual formula
%$p_{F_e} = \left(3\pi^2 n_e\right)^{1/3}$ and also the Debye mass
%$m_D^2=\frac{4\alpha}{\pi}\,p_{F_e}^2$.

Basically, in the  cold phase, which is expected to be realized in most regions of the NS  crust, ions form a Coulomb crystal with a body-centered cubic (bcc) lattice structure \cite{BarbaPhysRevC.106.065806}.
In an ideal crystal at zero temperature, $S(q)$ consists of discrete Bragg peaks at reciprocal lattice vectors while for warm crusts structures start to relax (see more recently~\cite{Barba-Gonzalez:2024xhc}). However, for transport and emissivity calculations it is customary to use an effective, angle-averaged static structure factor that accounts for phonon excitations and suppresses coherent scattering at small momentum transfer. A simplified and yet widely adopted parametrization is
\begin{equation}
S(q)=1-\exp\!\left[-(q a_i)^2\right], 
\end{equation}
where $a_i=\left(\frac{3}{4\pi n_Z}\right)^{1/3}$ is the ion Wigner--Seitz radius.
This form ensures that $S(q)\rightarrow 0$ for $q\rightarrow 0$, reflecting
the suppression of long-wavelength density fluctuations in a crystalline
lattice, while $S(q)\rightarrow 1$ for $q\gg a_i^{-1}$, where ion correlations become negligible. Together, these ingredients provide a realistic and self-consistent
description of ion correlations, nuclear finite-size effects, and electron
screening in the evaluation of the Coulomb logarithm entering the
electron--ion bremsstrahlung emissivity in neutron star crusts. 

%Finally, in differential form
%\begin{equation}
%\frac{d   Q_{a,eZ}}{d x}=  Q_{a,eZ} \Phi_{a,eZ}(x)=Q_{a,eZ}\frac{x^4}{24 \zeta(5)\left(e^x-1\right)}\,.
%\end{equation}
%\bjk{Same issue here - I think we should either use the $\Phi(x)$ notation or we should write $dQ/dx$ in full, but not both. }

\iffalse

\vspace*{0.5cm}

\item plasmon decay in magnetized plasma (or flares?). I don t know, to talk
\begin{equation}\left(\gamma^* \rightarrow a\right)
\end{equation}

For coupling to photons 
\begin{equation}
\mathcal{L} \supset-\frac{1}{4} g_{a \gamma} a F_{\mu \nu} \tilde{F}^{\mu \nu}
\end{equation}

one obtains a rate for $m_ a \ll \omega_p$
\begin{equation}
\Gamma_{\gamma^* \rightarrow a}=\frac{g_{a \gamma}^2}{96 \pi} \frac{\left(\omega_p^2-m_a^2\right)^3}{\omega_p^2}
\end{equation}

so that
$$
Q_{\gamma^* \rightarrow a}=\int \frac{d^3 k}{(2 \pi)^3} \omega(k) f_B(\omega) \Gamma_{\gamma^* \rightarrow a}
$$

where $f_B$ is the Bose-Einstein distribution for plasmons and in the limit $T \ll \omega_p$, only the mode $k \approx 0$ contributes

$$
Q \approx \frac{\omega_p^3}{2 \pi^2} \Gamma_{\gamma^* \rightarrow a} e^{-\omega_p / T} .
$$

with monoenergetic values

$$
E_a=\omega_p
$$
and finally

$$
\frac{d Q}{d E_a} \simeq Q_{\gamma^* \rightarrow a} \delta\left(E_a-\omega_p\right)
$$
\fi

\subsection{ALP photon coupling: Primakoff process on ions}

We consider thermal photon conversions in the electrostatic field of electrons and nuclei $$\gamma + Ze \rightarrow Ze + a\,.$$
The axion emissivity is 
\begin{equation}
Q_{a,\gamma Ze}=2 \int \frac{d^3 k}{(2\pi)^3} \Gamma_{\gamma \rightarrow a}(E) \frac{E}{e^{E/T}-1},
\end{equation}
where the factor of $2$ accounts for the photon polarizations and the denominator introduces the Bose-Einstein distribution function of the thermal photons. $k$ is the photon $3-$momentum and $\Gamma_{\gamma \rightarrow a}(E)$ is the rate of the process. Following ~\cite{Carenza:2020zil}, we work in the relativistic, negligible-recoil limit, assuming $m_a \ll E$ and $E_\gamma \simeq k \simeq E_a$.
Therefore, the differential emissivity in the parametrized form reads %\bjk{Do we have a reference for this? Also where did $\Gamma$ disappear to?}
\begin{equation}
\begin{aligned}
\frac{d Q_{\gamma Ze}}{d x}&=  \frac{g_{a \gamma}^2 T^5 \kappa_s^2}{32 \pi^3} \frac{x^3}{e^{x}-1}\left[\left(1+\frac{\kappa_s^2}{4 x^2 T^2}\right) \ln \left(1+\frac{4 x^2 T^2}{\kappa_s^2}\right)-1\right] \\ & \overset{x<<1}{\simeq} \frac{g_{a \gamma}^2 T^5 \kappa_s^2}{32 \pi^3} \frac{x^3}{e^{x}-1} \left(\frac{2 x^2 T^2}{\kappa_s^2}-\frac{8x^4T^4}{3\kappa^4_s}\right).
\label{diffprimakoff}
\end{aligned}
\end{equation}

In the case of neglecting degeneracy effects and plasma frequency, the charge screening can be well approximated by the Debye-Huckel screening wavenumber scale \cite{Carenza:2024ehj}
\begin{equation}
\kappa^2_s(r)=\frac{4\pi \alpha}{T}n_B(r)(Y_e(r)+ Z_j^2(r) Y_j(r)),
\end{equation}
where $Y_e$ and $Y_j$ are the electron number fraction per baryon and $j$th-nuclear species (charge $Z_je$ per baryon, respectively \cite{Pons2019}. %\bjk{This information about screening, plasma frequency etc comes out of nowhere. It's not clear why it's being said. Should it appear further down instead?}
Since $\kappa_s\gg E$, we perform an Taylor expansion of the logarithm in Eq.~(\ref{diffprimakoff}), written up to second order so as to include the number density contribution in-built from $\kappa_s$.

\begin{figure}[h!]
\centering
\includegraphics[width=0.495\linewidth]{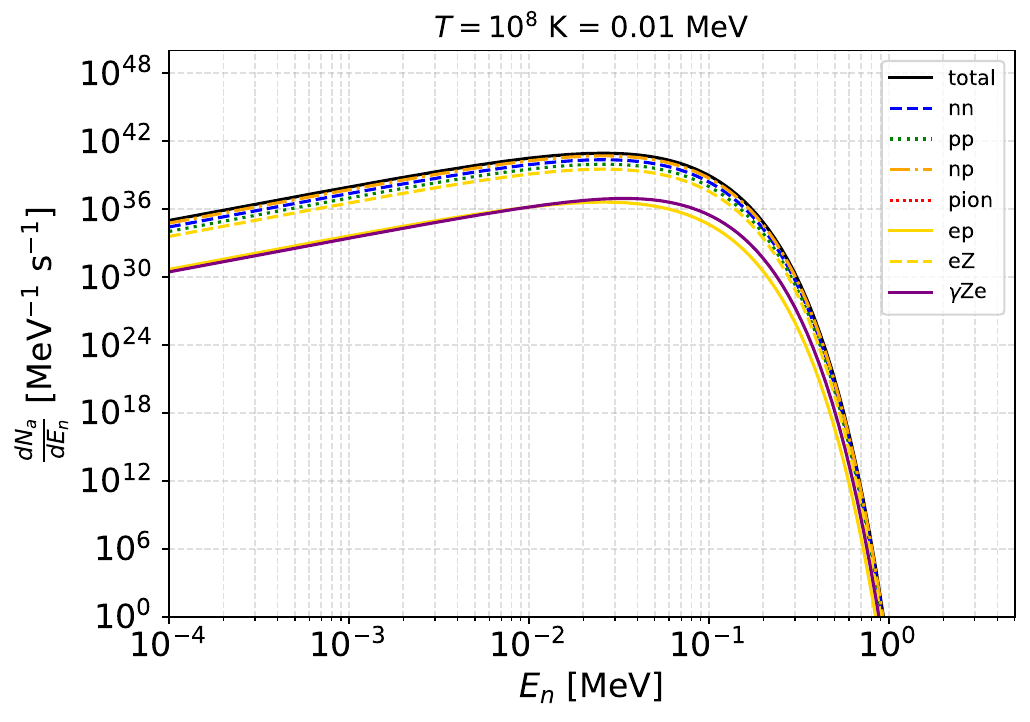}
\includegraphics[width=0.495\linewidth]{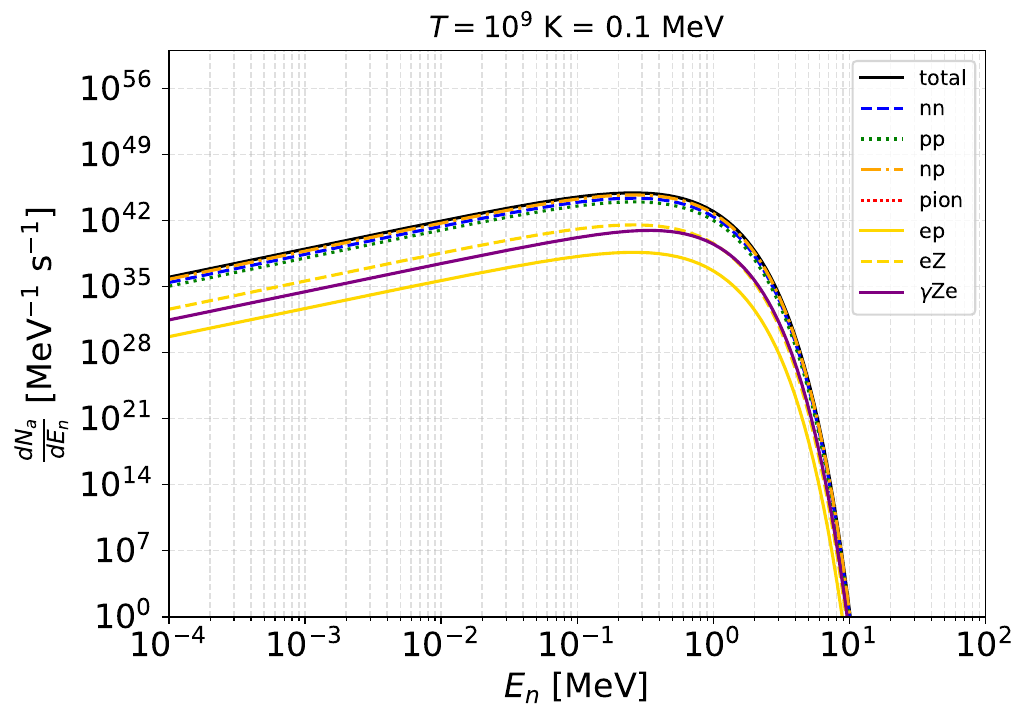} 
\includegraphics[width=0.495\linewidth]{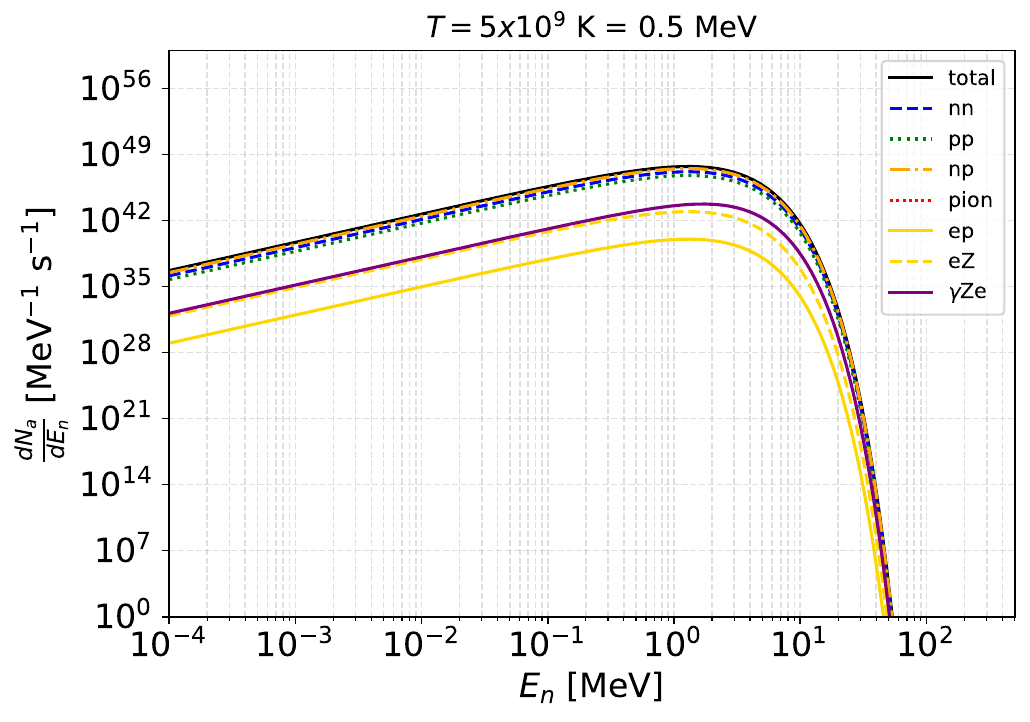} 
\includegraphics[width=0.495\linewidth]{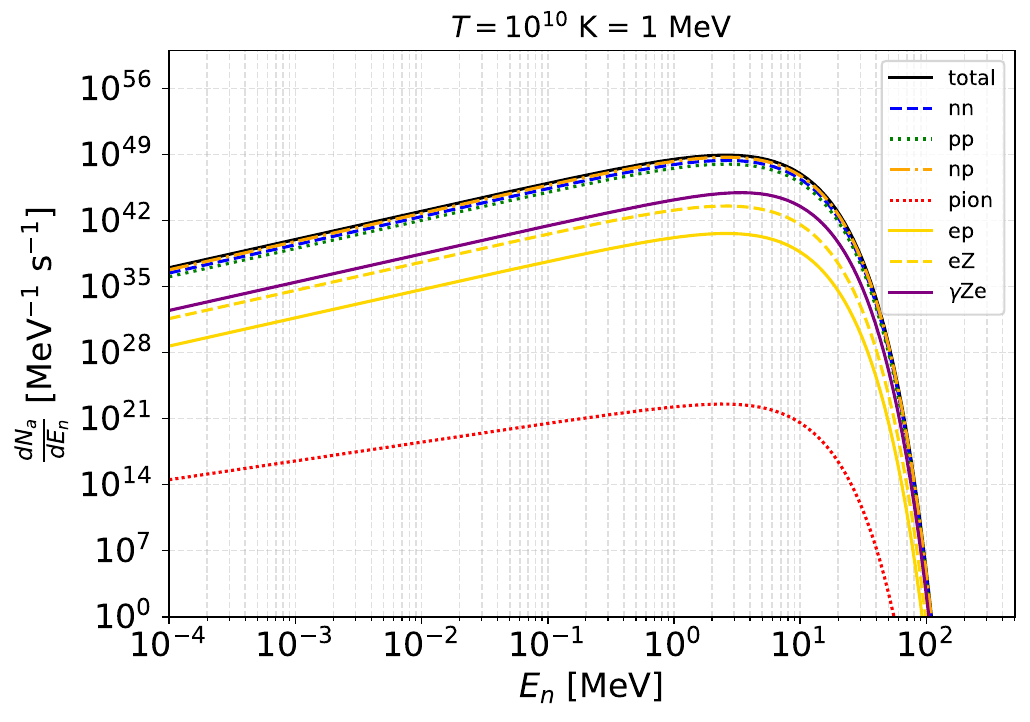} 
\caption{Energy spectra of ALPs $[\text{MeV}^{-1}\,\mathrm{s}^{-1}]$ produced for different internal NS temperatures: $T = 10^{8}\,\mathrm{K}$, $10^{9}\,\mathrm{K}$, $5\times10^{9}\,\mathrm{K}$, and $10^{10}\,\mathrm{K}$, showing  separately the contributions from the various production mechanisms. In each of these panels, the total is represented as a black solid line; the nucleon bremsstrahlung contributions from neutron--neutron, proton--proton, and neutron--proton interactions are shown by a blue dashed line, green dotted line, and orange dot--dashed line respectively. The pion--axion conversion contribution is represented by a red dotted line, while the ALP bremsstrahlung from electron--proton scattering is shown by a golden solid line and the ALP bremsstrahlung from electron--nuclei scattering (in the crust) is shown by a golden dashed line. We use couplings  $g_{an}=g_{ap}=g_{ae}=g_{a\gamma}=10^{-10}$ GeV$^{-1}$.}
\label{prod}
\end{figure}

\begin{figure}[h!]
    \centering

    \includegraphics[width=0.8\linewidth]{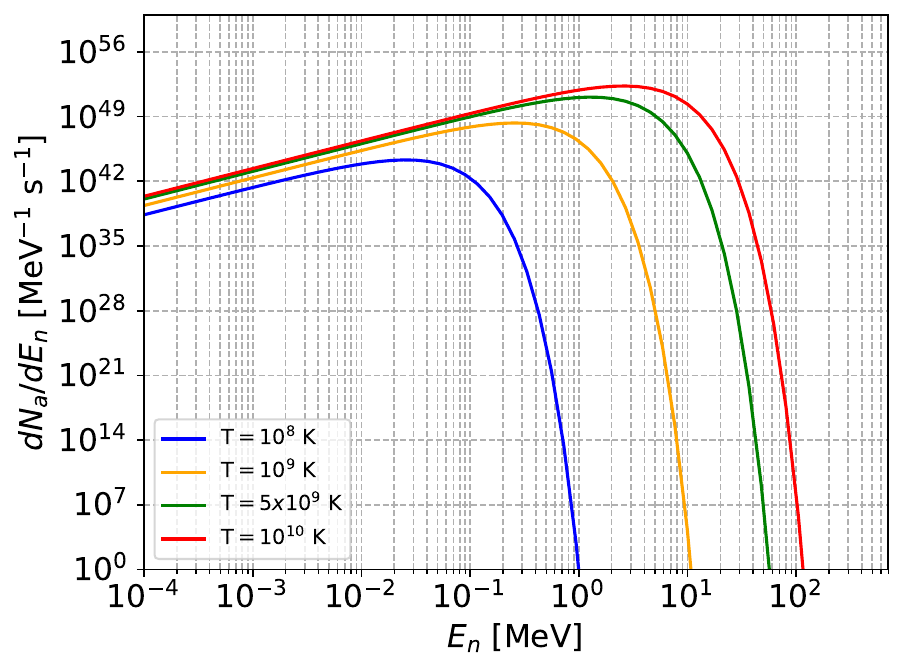}
    
    \caption{Total ALP energy spectra $[\,\mathrm{MeV^{-1}}\,s^{-1}]$ obtained by summing all individual production channels, 
    for the four temperatures considered (blue, red, green, and orange lines). $g_{ap}=g_{an}=5\times10^{-9}\text{ GeV}^{-1},\,g_{a\gamma}=10^{-11}\text{ GeV}^{-1},\,g_{ae}=1.3\times10^{-13}\text{ GeV}^{-1}$.}
    
    \label{fig:totalprod}
\end{figure}

\subsection{ALP Production Summary}

In Fig.~\ref{prod}, we show the ALP energy spectra %\footnote{For obtaining the dimensionless ALP energy spectrum,  we normalize $dN_a/dE$  by a factor$1.5\times10^{15}$, which arises from the typical NS  volume and the conversion between energy and number emission rates in cgs units. \bjk{I'm a bit confused by this. Isn't all this normalisation already included in the calculation of $dN/dE$ (since that includes an integral over the volume)? Also I don't think we need to specify the unit conversion factors (since we explicitly give the units in the plot.}} 
for four representative values of the internal temperature in the range $T\in[10^8,\,10^{10}]\,\mathrm{K}$, assuming benchmark  ALP couplings $g_{an}=g_{ap}=g_{ae}=g_{a\gamma}=10^{-10} \, \rm GeV^{-1}$. 
At high energies the ALP spectra rapidly decrease and tend to zero, so that only at sufficiently high temperatures could they contribute significantly at the Fermi-LAT threshold of $E \sim 100~\mathrm{MeV}$. The ALP spectrum is strongly suppressed at large energies due to the exponential factor $\sim e^{-x}$ appearing in the thermal distribution functions above. 
Conversely, at energies $\lesssim 1~\mathrm{MeV}$ the ALP production becomes efficient even for relatively low temperatures. At all temperatures studied, the main contribution comes from nucleon-nucleon bremsstrahlung. 
Therefore, we conclude that when considering astrophysical objects characterized by such temperatures, it is necessary to rely on soft X-ray experiments to probe the corresponding ALP emission. In Fig.~\ref{fig:totalprod}, we show the total ALP production rate for each temperature using alternative benchmark values motivated by current limits $g_{ap}=g_{an}=5\times10^{-9}\text{ GeV}^{-1},\,g_{a\gamma}=10^{-11}\text{ GeV}^{-1},\,g_{ae}=1.3\times10^{-13}\text{ GeV}^{-1}$~\cite{OHare2020}. In the following sections, we will fix the couplings $g_{an},\,g_{ap},\,g_{ae}$ to these values, and our analysis will focus on varying $g_{a\gamma}$.

%------------------------------------------------------------
% Semi-Compton (Compton-like) production on relativistic, degenerate electrons
% Process:  gamma^{(*)}(k) + e^-(p) -> a(q) + e^-(p')
%------------------------------------------------------------
\vspace*{0.2cm}
\section{ALP conversion}
\label{section:lpconversion}

In this section we assume that ALPs produced in the NS core can escape to the more external layers, such as the NS atmosphere and magnetosphere, where they can convert into photons while propagating in the existing magnetic fields due to the non-vanishing ALP-photon coupling. In addition, propagation to Earth in the galactic magnetic field plays a crucial role that we will address later in this section.

\subsection{Conversion inside the NS magnetosphere}\label{conversioninside}

In this section, we aim to compute the probability of ALP conversion in the NS magnetosphere. Given the significant intrinsic uncertainties associated with the stellar magnetic field structure and the particle density distribution, we adopt a simplified description, assuming a dipolar magnetic field %(Subsec. \ref{conversioninside}) 
\begin{equation}
    B(r,\theta)=B_0\left(\frac{r_0}{r}\right)^3 (2 \cos\theta \,\hat{r}+\sin\theta \,\hat{\theta})\,,
\end{equation}
where $\theta$ is the colatitude angle of the dipolar magnetic field. The photon–ALP mixing is sourced by the component of the magnetic field transverse to the ALP propagation direction $\hat{k}$, namely
\begin{equation}
    B_T = |\mathbf{B} \times \hat{k}|\,.
\end{equation}
For simplicity, we assume that the ALPs propagate radially outward from the NS toward the magnetosphere. This approximation allows us to describe the conversion in the adopted dipolar magnetic-field geometry while neglecting the additional geometric effects associated with non-radial trajectories.\footnote{See e.g.~Refs.~\cite{Witte:2021arp,McDonald:2023shx,Gines:2024ekm} for a more detailed and general treatment.} For radial ALP propagation ($\hat{k}=\hat{r}$), only the $\hat{\theta}$ component contributes, yielding
\begin{equation}
    B_T = B_0 \left(\frac{r_0}{r}\right)^3 \sin\theta\,,
\end{equation}
where \(r_0\equiv R\) denotes the stellar radius and \(r\geq r_0\). For simplicity, we consider an equatorial trajectory ($\theta=\pi/2$), maximizing $B_T$, while the general case can be recovered by including an overall $\sin^2\theta$ factor \cite{Fiorillo:2025gnd}.

Since we are interested in the ALP regime $E \gg m_a$ the short-wavelength approximation can
be applied successfully \cite{Raffelt:1987im} %and turns the beam propagation equation into the 
to obtain the dynamical equation in the Schrodinger-like form
\begin{equation}
    i\,\frac{d}{dr}
\begin{pmatrix}
A_{\perp}(r) \\[4pt]
A_{\parallel}(r) \\[4pt]
a(r)
\end{pmatrix}
= 
\left(E + \mathcal{M}\right)
\begin{pmatrix}
A_{\perp}(r) \\[4pt]
A_{\parallel}(r) \\[4pt]
a(y)
\end{pmatrix}\,,
\end{equation}
where the ALP/photon propagation is radial, with $r$ being the dimensional distance from the NS core. $A_{\parallel}(r)$ and $A_{\perp}(r)$ are the two photon linear polarization amplitudes parallel and perpendicular to the transverse magnetic field $B_T=B(r)\,\sin{\theta}$ with respect to the propagation direction, respectively. The ALP amplitude is denoted by $a(r)$, and $\mathcal{M}$ represents the photon-ALP mixing matrix
\begin{equation}
    \mathcal{M} =
\begin{pmatrix}
\Delta_{\perp} & 0 & 0 \\[6pt]
0 & \Delta_{\parallel} & \Delta_{a\gamma} \\[6pt]
0 & \Delta_{a\gamma} & \Delta_{a}
\end{pmatrix}\,,
\end{equation}
where $\Delta_{a}=-\frac{m_a^2}{2E_n}$ is the ALP mass term,
$\Delta_{a\gamma}=\frac{1}{2}g_{a\gamma\gamma} B \sin\theta\,$ is the off-diagonal ALP-photon mixing term, with $\theta$ angle between propagating ALP/photon direction and $B$ and
$\Delta_{\parallel, \perp}=\Delta_{pl}+\Delta_{QED,\parallel, \perp}=-\frac{\omega_{pl}^2}{2E_n}+\Delta_{QED, \parallel, \perp}\,$ is the photon diagonal term.\footnote{We have considered only the $\Delta_{\parallel}$ term, since only the linear polarization component of the photon that is parallel to $B_T$ can mix with the ALP field.} The plasma frequency is given by $\omega^2_{pl}=\frac{4\pi\alpha\, n_e}{m_e}$. %\textcolor{red}{We need to decide the value of $n_{e,0}$.} 

%In the case of very energetic transitory events featured by magnetars, the charge density near the star surface is $n_e \sim 10^{30} \textit{cm}^{-3}$ \cite{DeMiguel:2022ojb}, while in steady regime the outer layer of NS has a $n_{e,0}\simeq 10^{25}cm^{-3}$ decreasing in radial way in the magnetosphere \footnote{In addition, the EBL and starlight may have profound effects on the refractive index at large energies: when $E_{CMB}<400\,TeV$, a $\Delta_{CMB}\simeq 0.522\times10^{-42}E_{\gamma}$ has to take into account.}.

The Goldreich–Julian (GJ) model \cite{Goldreich:1969sb} describes a plasma-filled neutron star magnetosphere rotating with angular frequency $\Omega=2\pi/P$ (rotation period $P$), permeated by a corotating dipolar magnetic field in flat spacetime, and provides analytic expressions for the charge density and magnetic field structure. It is typically adopted as a good approximation for the closed-field regions of the magnetosphere ($\lesssim 10^9$ cm). The GJ charge density is:
\begin{equation}
n_e(r)=\kappa(r)n_{GJ}(r)
=\kappa(r)\frac{\Omega B_0}{2\pi ec}
\left(\frac{r_0}{r}\right)^3
=\kappa(r)\frac{B_0}{ec\,P}
\left(\frac{r_0}{r}\right)^3.
\end{equation}
with $10^2\leq \kappa \leq 10^5$ representing a secondary electron-positron pair multiplicity factor. In realistic pulsar magnetospheres, the primary particles accelerated in the polar cap trigger a cascading production of $e^\pm$ pairs through curvature radiation and subsequent photon-photon absorption in the intense magnetic field, scaling the local plasma density well above the baseline Goldreich--Julian value \cite{Goldreich:1969sb, Timokhin:2015dua}. For example, for PSR J1119$-$6127, a magnetar-like pulsar, we expect
%$n_{e,0}= \kappa\times B_0/(ecP)\sim (10^4)\cdot (5\times10^{13}\,\mathrm{G})/\left[(4.803\times10^{-10})(2.998\times10^{10})(409\,\mathrm{ms})\right]=8.5\times10^{16}\,\mathrm{cm}^{-3}$. \bjk{<- The values $B_0$ etc are given in Table 1, so maybe we don't need to write out all these numbers explicitly. Maybe we can just write "
$n_{e,0} \sim \kappa\times B_0/(ecP) = 7\times10^{16}\,\mathrm{cm}^{-3}$, where we have assumed $\kappa \sim 10^4$. See the Table \eqref{tab:pulsars_transposed_ul} for the $n_{e,0}$ value of the other PSRs.

Very close to the NS surface, however, the
GJ model does not accurately capture the electron density. Indeed, values of the internal number density of electrons we expect at the surface are almost $10^{35}\,\mathrm{cm}^{-3}$ (as we have described at the beginning of Sec. \ref{section:ALPproduction}). This value arises from a steep decrease in the atmosphere layer from crust values. Basically, in the outermost layers of a neutron star, the atmosphere forms an extremely thin, highly compressed plasma layer characterized by a steep density gradient.
Assuming hydrostatic equilibrium and an isothermal atmosphere it  follows an exponential profile, $\rho_{\rm atm}(z) = \rho_0\, e^{-z/H}$, where $z = r - R$ denotes the height above the stellar surface, $\rho_0$ is the density at the base of the atmosphere, and $H$ is the density scale height. The corresponding electron number density can be written as
\begin{equation}
n_{e, \rm atm}(z) \simeq  6.0 \times 10^{23}\, Y_e
\left(\frac{\rho_0}{1\,\mathrm{g\,cm^{-3}}}\right)
e^{-z/H}\ \mathrm{cm^{-3}}.
\end{equation}
%where $Y_e = Z/A$ is the electron fraction.
This atmosphere versus magnetosphere hierarchy is consistent with standard  NS atmosphere models and highlights the sharp transition between the dense atmospheric plasma and the dilute magnetospheric environment. The density scale height can be estimated from hydrostatic equilibrium as
\begin{equation}
H \simeq \frac{k_B T}{\mu m_u g_*},
\end{equation}
where $m_u$ is the atomic mass unit, $\mu$ is the mean molecular weight, and $g_*$ is the surface gravitational acceleration 
\begin{equation}
g_* \simeq 1.9 \times 10^{14}
\left(\frac{M}{1.4\,M_\odot}\right)
\left(\frac{R}{10\,\mathrm{km}}\right)^{-2}
\left(1 - \frac{2GM}{R c^2}\right)^{-1/2}
\ \mathrm{cm\,s^{-2}}.
\end{equation} For typical values, this yields $H \lesssim 1 - 10\ \mathrm{cm}$. %\bjk{I think I'm a little confused by the discussion starting from "But looking at the values of internal number density of electron..." up to here. I think maybe there just needs to be an extra sentence to sign-post where the discussion is going. For example, "Very close to the NS surface, however, the GJ model does not accurately capture the electron density." (followed by the discussion of the atmosphere).}

The vacuum refraction parameter of quantum electrodynamics (QED), $\Delta_{QED,\parallel,\perp}$, comes from the Euler-Heisenberg Lagrangian term \cite{Heisenberg:1936nmg, Weisskopf:1936, Schwinger:1951nm}:
\begin{equation} {\cal L}_{\rm HEW} = \frac{\alpha^2}{90 m^4_e}\left[(F_{\mu\nu}F^{\mu\nu})^2+\frac{7}{4}(F_{\mu\nu}\tilde{F}^{\mu\nu})^2\right],\end{equation}
which takes into account the effects of vacuum polarization below the threshold for electron-positron pairs. According to Adler's  theory \cite{Adler:1971wn}, its contribution is given by the exact one-loop polarization function,
\begin{equation}
\Delta_{QED,\parallel,\perp}=\frac{1}{2E_n}\,\Pi_{\parallel,\perp}(B,E_n)\,.
\end{equation}
The generalized longitudinal polarization function in a constant magnetic field is
\begin{equation}
\Pi_\parallel(B,E_n)=
\frac{\alpha}{2\pi}\!\int_0^\infty\!\frac{ds}{s^{2}}\,e^{-m_e^2 s}
\!\left[(eBs)\coth(eBs)-1-\frac{(eBs)^2}{3}\right]
-
\frac{\alpha\,E_n^2\sin^2\theta}{3\pi}
\int_0^\infty ds\,e^{-m_e^2 s}
\left[1-(eBs)\coth(eBs)\right]\,.
\end{equation}
The corresponding transverse eigenvalue of the polarization tensor, which does not mix with the axion is
\begin{equation}
\Pi_\perp(B,E_n)=
\frac{\alpha}{2\pi}\!\int_0^\infty\!\frac{ds}{s^{2}}\,e^{-m_e^2 s}
\!\left[(eBs)\coth(eBs)-1-\frac{(eBs)^2}{3}\right]
+
\frac{\alpha\,E_n^2\sin^2\theta}{6\pi}
\int_0^\infty ds\,e^{-m_e^2 s}
\left[1-(eBs)\coth(eBs)\right]\,.
\end{equation}
Thus the photon diagonal terms are
\begin{equation}
\Delta_\parallel=-\frac{\omega_{\rm pl}^2}{2E_n}+\frac{1}{2E_n}\Pi_\parallel(B,E_n),
\qquad
\Delta_\perp=-\frac{\omega_{\rm pl}^2}{2E_n}+\frac{1}{2E_n}\Pi_\perp(B,E_n).
\end{equation}

In the weak-field limit \(B\ll B_\mathrm{cr}\) (with $B_\mathrm{cr}\simeq 4.414\times 10^{13}$ G critical magnetic field), valid at large radii, the Euler–Heisenberg expansion is recovered.  One finds
\begin{equation}
\Pi_\parallel \simeq \frac{7\alpha}{45\pi}\,E_n^2\left(\frac{B}{B_\mathrm{cr}}\right)^{2},
\qquad
\Pi_\perp \simeq \frac{4\alpha}{45\pi}\,E_n^2\left(\frac{B}{B_\mathrm{cr}}\right)^{2}\,.
\end{equation}
These expressions describe vacuum birefringence in the perturbative regime.

In the strong-field limit \(B\gg B_\mathrm{cr}\), close to the surface of astrophysical objects such as magnetars, the polarization functions become independent of the photon energy and scale linearly with the magnetic field strength,
\begin{equation}
\Pi_\parallel \simeq \frac{\alpha}{2\pi}\,eB,
\qquad
\Pi_\perp \simeq \frac{\alpha}{6\pi}\,eB.
\end{equation}
In this regime the longitudinal mode acquires a significantly larger effective mass than the transverse mode, preventing axion–photon mixing close to the surface of strongly magnetized neutron stars.  Mixing becomes efficient only at larger distances, where the magnetic field has decreased sufficiently for the system to re-enter the weak-field Euler–Heisenberg regime. In the strong regime, with \(B\gg B_\mathrm{cr}\), since the plasma and QED contributions share the same dependence on energy and radius, the QED term is always larger than the plasma term.

\begin{figure}[h!]
\centering
\includegraphics[width=0.495\linewidth]{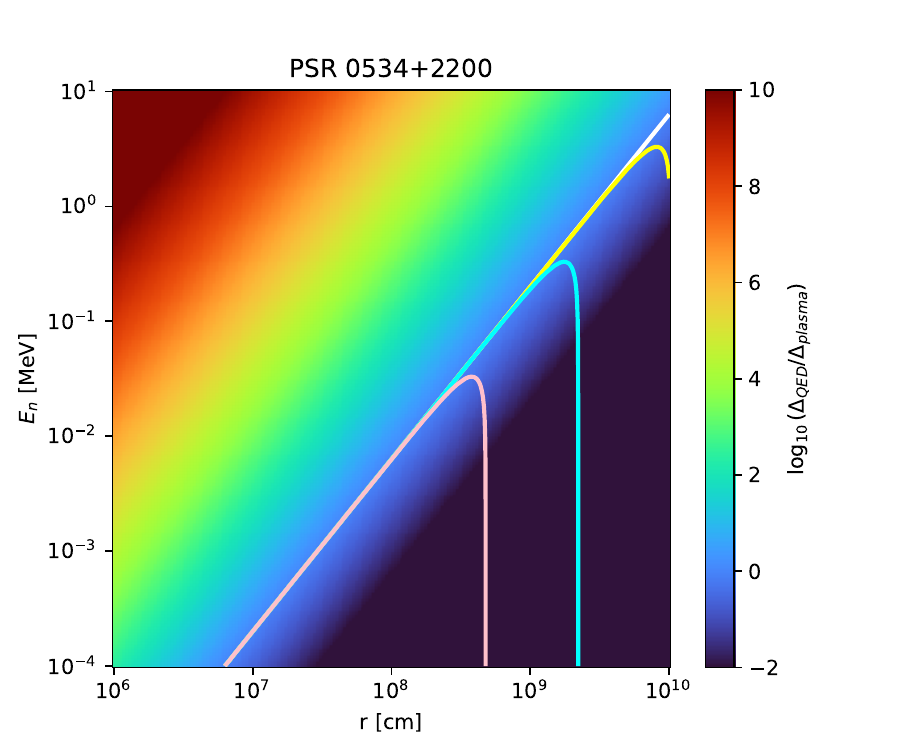} 
\includegraphics[width=0.495\linewidth]{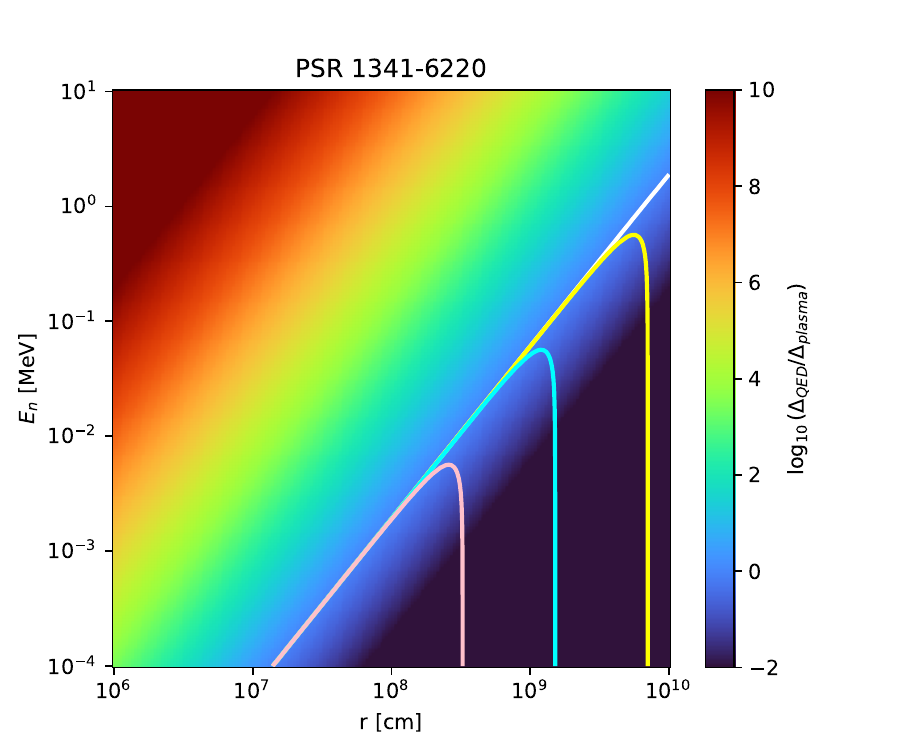} 
\includegraphics[width=0.495\linewidth]{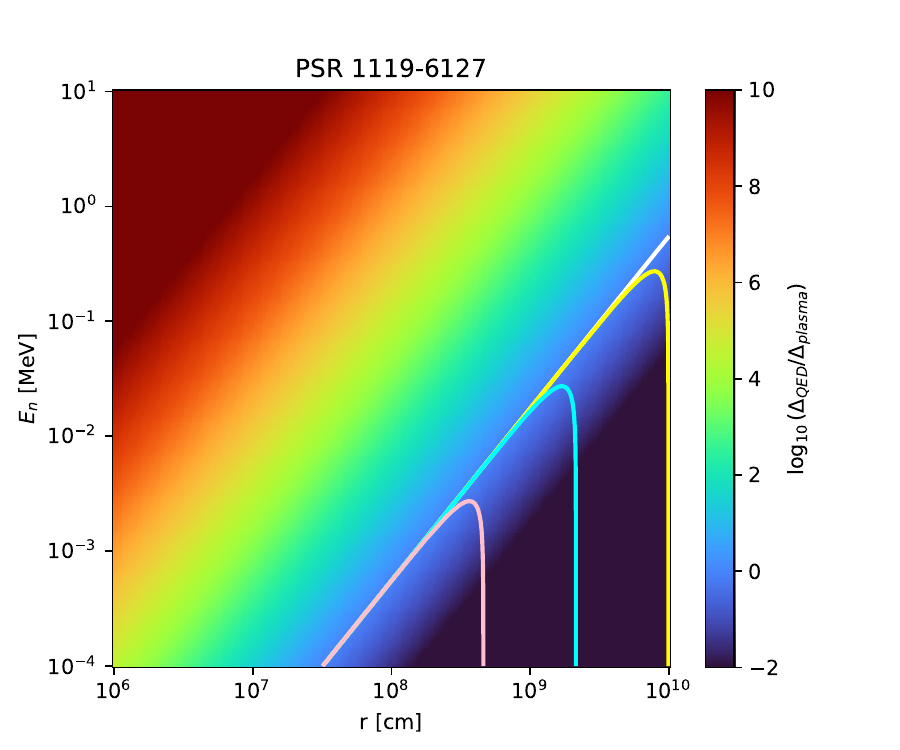} 
\includegraphics[width=0.495\linewidth]{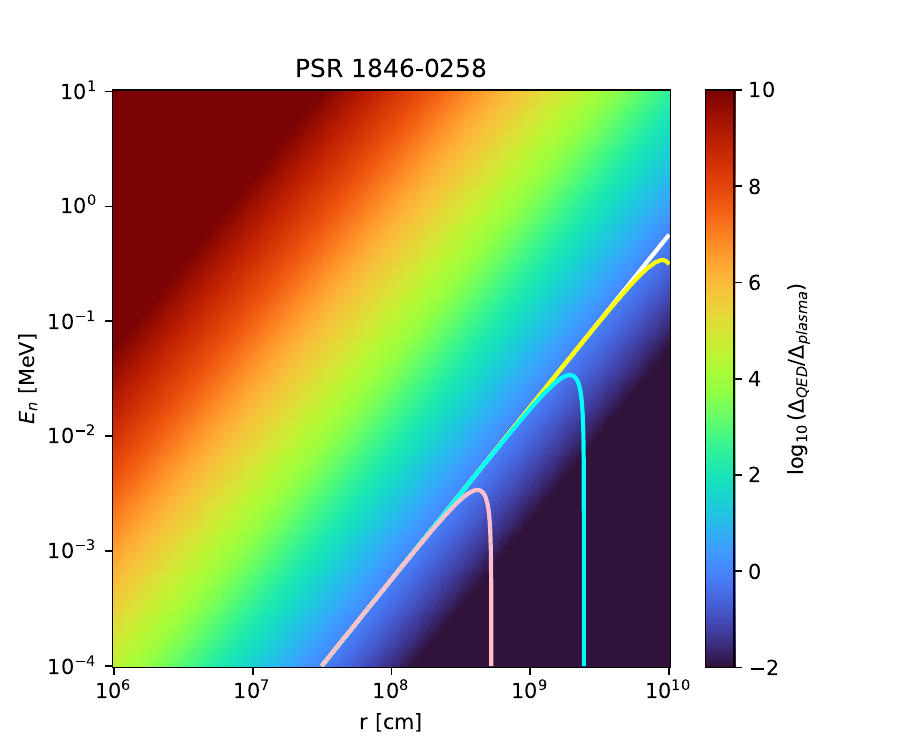} 
    \caption{Parameter space of ALP energy $E$ vs radius $r$ for four PSRs ($B_0$ and $n_{e,0}$ of each PSR are shown in Tab.\ref{tab:pulsars_transposed_ul}). The color gradient is the logarithmic ratio of the QED term over the plasma term. The solid lines mark the resonance condition of Eq.~\eqref{eq:resonance}. The ALP mass is fixed at $m_a=10^{-9}$ eV (white), $m_a=10^{-8}$ eV (yellow), $m_a=10^{-7}$ eV (cyan), $m_a=10^{-6}$ eV (pink).}
    \label{ratio}
\end{figure}

In Fig.~\ref{ratio} we show the $(E, r)$ parameter space of the PSRs that we will analyze, indicating the regimes in which the relative importance of the two $\Delta$ terms, QED and plasma, changes, and with one term dominating over the other. In Fig.~\ref{ratio}, in the region of interest in terms of both energy and radial distance, the plasma term is negligible and the QED term dominates the diagonal component of the ALP-photon mixing and therefore governs the conversion (red-orange-green region), while in the blue region both contributions are comparable. In the outer magnetosphere (dark region), plasma effects contribute significantly to the ALP–photon conversion dynamics. For the distance chosen, we have considered only the weak-field limit.

The conversion is maximal under the resonance condition, which occurs when the difference between the photon and ALP dispersion relations vanishes, including the QED polarization term:
\begin{equation}
\label{eq:resonance}
    \Delta_{\parallel}-\Delta_a\simeq 0 \quad \Rightarrow \quad \frac{m^2_a-\omega^2_{pl}}{2E_n}+\frac{7\alpha}{90\pi}\left(\frac{B}{B_\mathrm{cr}}\right)^2 E_n\simeq 0\,.
\end{equation} 

%quello che si vede è che il termine QED fa si che la condizione di risonanza dipenda da E^2: se elimino il termine QED si ha una linea verticale perchè si verifica solo a una determinata distanza. Però succede che aumentando la massa si arriva a un punto in cui QED è cosi piccolo, rispetto a ma che si trascura e la massa ma matcha perfettamente la densità di plasma.

As visible from the white, yellow, cyan and pink lines in Fig.~\ref{ratio}, this resonance condition traces the cross-over region where the plasma and mass mismatch is non-trivially compensated by the QED vacuum polarization term when there is energy dependence, while once the ALP mass increases the QED term is negligible and the position of resonance is energy independent. 

We define the strong mixing regime within the energy window $[E_\mathrm{cr}, E_\mathrm{max}]$, where $E_\mathrm{cr}=\frac{|m^2_a-\omega^2_{pl}|}{2 g_{a\gamma}B}$ marks the onset of large oscillations and $E_\mathrm{max}=\frac{90\pi}{7\alpha}\frac{B^2_{cr}g_{a\gamma}}{B}$ limits its validity due to the QED vacuum polarization ($\Delta_{QED}$). $E_\mathrm{cr}$ exceeds $E_\mathrm{max}$ in certain regions of the parameter space, completely suppressing the conversion. Near the stellar surface ($r=1.5r_0$), the immense magnetic field enhances $\Delta_{QED}$, forcing $E_\mathrm{max}$ below $E_\mathrm{cr}$. Conversely, at larger distances ($r\geq 10^3\,r_0$), the dipolar decay of the magnetic field ($B \propto r^{-3}$) relaxes this QED buffering effect. Since $E_\mathrm{max} \propto 1/B$, the energy window widens significantly. For instance, fixing $g_{a\gamma}=10^{-11}$ GeV$^{-1}$ and $m_a=10^{-9}$ eV, strong mixing is forbidden close to the star but becomes allowed for $r=10^3\,r_0$ ($E_\mathrm{cr}=9\times10^{-4}$ MeV, $E_\mathrm{max}=5.11\times10^{-2}$ MeV), expanding by a factor of 1000 at $r=10^4\,r_0$ where $E_\mathrm{max}$ reaches $51.1$ MeV.

We define a quantity related to the oscillation length $l_{osc}\simeq 2\pi/\Delta_{osc}$ as:
\begin{equation}
    \Delta^2_{osc}=(\Delta_{\parallel}-\Delta_a)^2+4\Delta_{a\gamma}^2=\left(-\frac{\omega^2_{pl}}{2E_n}+\frac{7}{2}\frac{\alpha E_n}{45\pi}\left(\frac{B}{B_\mathrm{cr}}\right)^2+\frac{m^2_a}{2E_n}\right)^2+\left(g_{a\gamma\gamma}B \sin{\theta}\right)^2\,.
    \label{eq:oscillation-length}
\end{equation}
This length is related to the frequency of ALP-photon oscillation conversion. If $l_{osc}$ is small the oscillations are rapid, otherwise we have wide oscillations. 

The conversion probability can be obtained by perturbation theory through the following expression \cite{Fortin:2021sst,Fiorillo:2025gnd}:
\begin{equation}
    P\left(a \rightarrow \gamma_{\|}\right)=\left| \mathcal{A}_{a\gamma}\right|^2= \left|\int_0^z d z^{\prime} \Delta_{a\gamma}\left(z^{\prime}\right) \exp \left\{\mathrm{i} \Delta_a z^{\prime}-\mathrm{i} \int_0^{z^{\prime}} (\Delta_{QED,\parallel}+\Delta_{pl})\left(z^{\prime \prime}\right) d z^{\prime \prime}\right\}\right|^2\,.
    \label{probabilityconversion}
\end{equation}
%
%Here $\Delta_{a\gamma}$ is the conversion strength, while the exponential represents the relative phase ALP-photon. In the paper by Fiorillo and Vitagliano \cite{Fiorillo:2025gnd}, they consider both the massless and massive axion cases while neglecting the resonant regime, since the plasma frequency is already small in the regions where the conversion is effective. Conversion is effective when the oscillation phase term is approximately of order one, while if this phase is much larger than one, the rapid oscillations suppress the conversion. Therefore, in the paper, the conversion radius is determined solely by the QED term in the massless case, and by the combination of QED and the axion mass term in the massive case. In the general case, to obtain the conversion radius, one must solve
%
Here $\Delta_{a\gamma}$ determines the strength of conversion, while the exponential represents the relative ALP-photon phase. In Ref.~\cite{Fiorillo:2025gnd}, they consider both the massless and massive axion cases while neglecting the resonant regime, since the plasma frequency is already small in the regions where the conversion is effective. 

Conversion is efficient when the oscillation phase remains coherent over the propagation distance, i.e., when the accumulated phase difference over a characteristic macroscopic scale $R$ is of order one:
\begin{equation}
    |\Delta_{\rm tot}(R)|\, R=|\Delta_{\rm QED}(R) + \Delta_{\rm pl}(R) - \Delta_a(R)| \, R = 1 \,.
    \label{conv_radius}
\end{equation}
Physically, this condition determines the \textit{conversion radius} $R_{\rm conv}$ by comparing the propagation distance with the local \textit{coherence length}, defined as
\begin{equation}
L_{\rm coh} \sim \frac{2\pi}{|\Delta_{\rm tot}(R)|}\,,
\end{equation}
which is the distance over which the ALP and photon remain in phase and conversion can build up. If the phase difference grows much larger than one, rapid oscillations lead to destructive interference and suppress the conversion. 

Therefore, the conversion radius represents the spatial boundary where the local coherence length becomes macroscopic ($L_{\rm coh} \sim R$), allowing the conversion to successfully activate. In the massless ALP case, $R_{\rm conv}$ is determined almost solely by the QED term due to its steep radial dependence ($\Delta_{\rm QED} \propto R^{-6}$), leading to the well-known scaling $R_{\rm conv} \propto B_0^{2/5}$. Conversely, in the massive case, the conversion radius arises from the combination of the QED birefringence and the constant axion mass term $\Delta_a$.

\begin{figure}[htbp!]
    \centering
    % Prima immagine
    \begin{minipage}{1\textwidth}
        \centering
        \includegraphics[width=\textwidth]{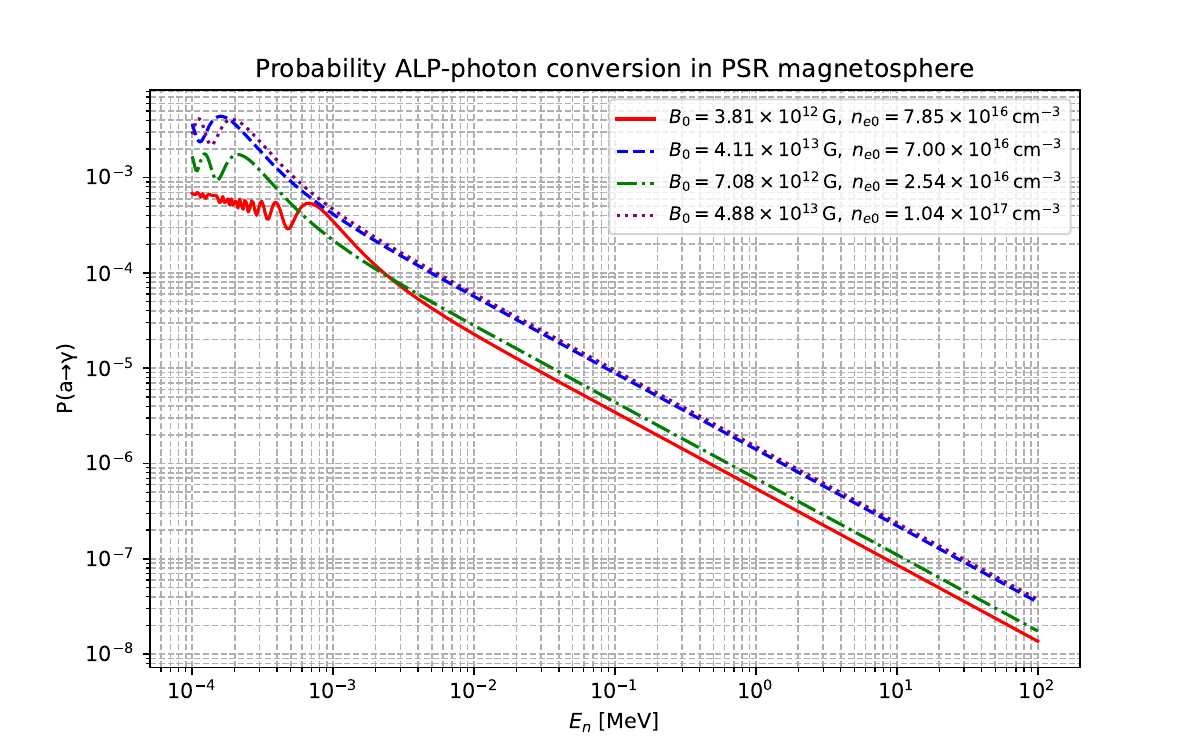}
    \end{minipage}
    \begin{minipage}{1\textwidth}
        \centering
        \includegraphics[width=\textwidth]{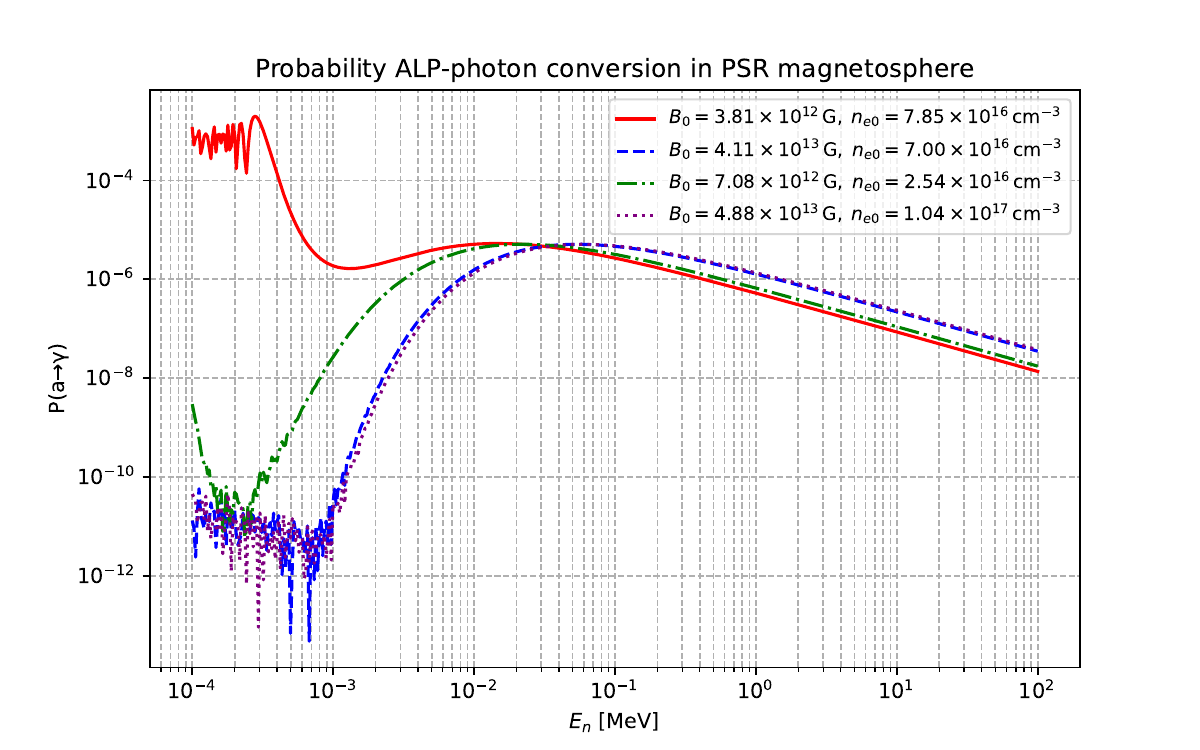}
    \end{minipage}
    \caption{Probability of conversion ALP into photons in PSR 1119-6127, PSR 1341-6220, PSR 0534+2200, PSR 1846-0258 magnetospheres for ALP mass $m_a=10^{-9}$ eV (\emph{top}) and $m_a=10^{-4}$ eV (\emph{bottom}), and ALP-photon coupling $g_{a\gamma}=10^{-11}$ GeV$^{-1}$.}
    \label{fig:convprobmagnetospheregag11}
\end{figure}

The conversion probability inside the magnetosphere (up to $r=10^4\,r_0$) of the four representative PSRs is shown in Fig. \ref{fig:convprobmagnetospheregag11}, for two values of ALP mass $m_a= 10^{-9}$ eV and $m_a= 10^{-4}$ eV, and a given value of ALP-photon couplings $g_{a\gamma}=10^{-11}$ GeV$^{-1}$. In this energetic regime, the conversion probability becomes completely independent of the ALP mass when it is very low. This occurs because the mass dispersion term $\Delta_a$ scales inversely with energy and it is entirely overwhelmed by the plasma frequency and the QED vacuum term throughout the integration path. The system thus operates in an effective massless particle regime, where the dynamics are dictated solely by the magnetic field geometry and the coupling $g_{a\gamma}$.

\subsection{Conversion in galactic magnetic field}\label{conversiongal}

We now consider the ALP fraction that remains unconverted in the NS magnetosphere and its possible subsequent conversion during propagation to Earth in the Galactic magnetic field $B_\text{galactic}$.

We compute this conversion process inside the Milky Way following the procedure outlined in \cite{2019MNRAS.487..123G}, adopting an updated description of the Galactic environment. The propagation is governed by the Galactic magnetic field $B_{\rm gal}$, whose strength is of order ${\cal O}(1) \, \mu \rm G$ and which consists of both regular and turbulent components. Since the $\gamma \leftrightarrow a$ oscillation length largely exceeds the coherence length of the turbulent field, only the regular component significantly affects the conversion probability.

For this component we use the model by Jansson and Farrar \cite{2012ApJ...757...14J,2012ApJ...761L..11J}, featuring disk and halo contributions parallel to the Galactic plane together with a central poloidal X-shaped structure. Its updated version \cite{2017ICRC...35..558U} incorporates improved polarized synchrotron data and refined cosmic-ray and thermal electron distributions. The turbulent field is described according to \cite{2016JCAP...05..056B}, although its impact is subdominant in the present context.

As a cross-check, we have also implemented the model by Pshirkov et al. \cite{2011ApJ...738..192P}, finding only minor quantitative differences. Nevertheless, we adopt the Jansson–Farrar model as our reference, since the Pshirkov et al. framework is mainly constrained by data along the Galactic plane and provides a less accurate description of the halo component.

The plasma contribution to the mixing matrix is determined by the Galactic electron density $n_{e,\, \rm gal}$. Inside the disk we take $n_{e,\, \rm gal} \simeq 1.1 \times 10^{-2} \, \rm cm^{-3}$, as obtained from the model of \cite{2017ApJ...835...29Y}, corresponding to a plasma frequency $\omega_{pl, \, \rm gal} \simeq 3.9 \times 10^{-12} \, \rm eV$.
This electron-density model includes the thick and thin disks, spiral arms, Galactic Center component, and relevant local structures, with parameters fitted to pulsar dispersion measures.

\begin{figure}[h!]
    \centering
\includegraphics[width=0.495\textwidth]{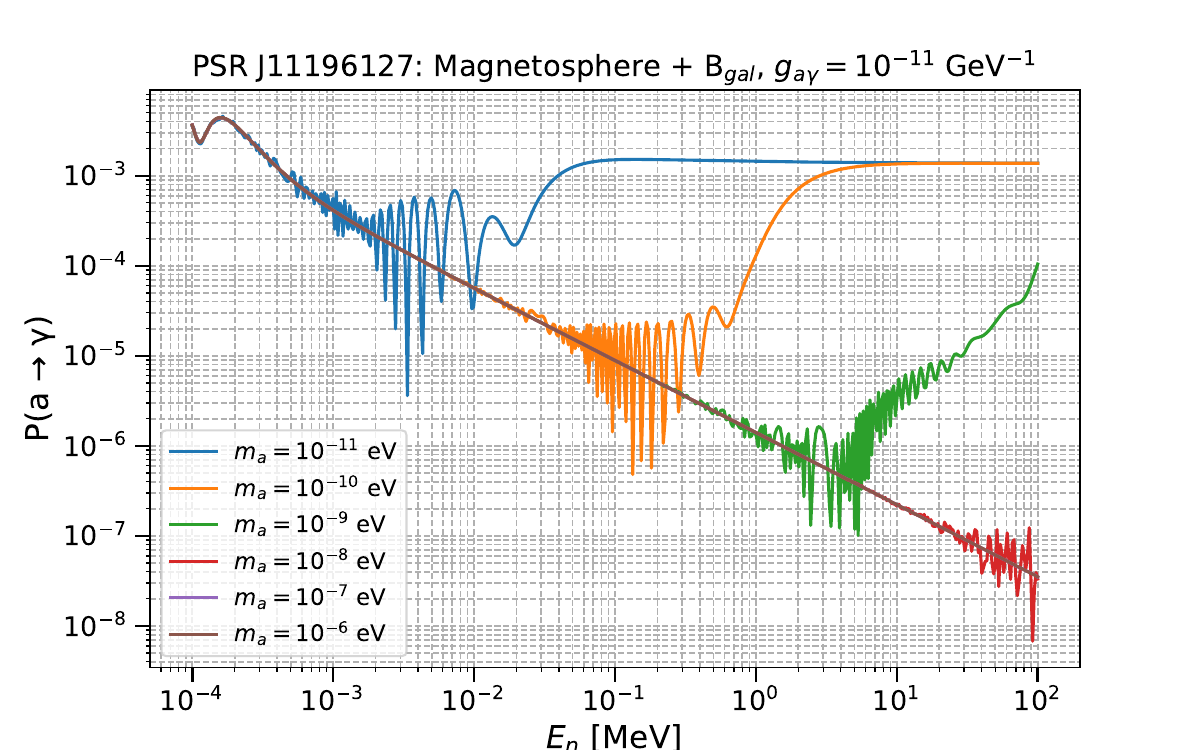}
\includegraphics[width=0.495\textwidth]{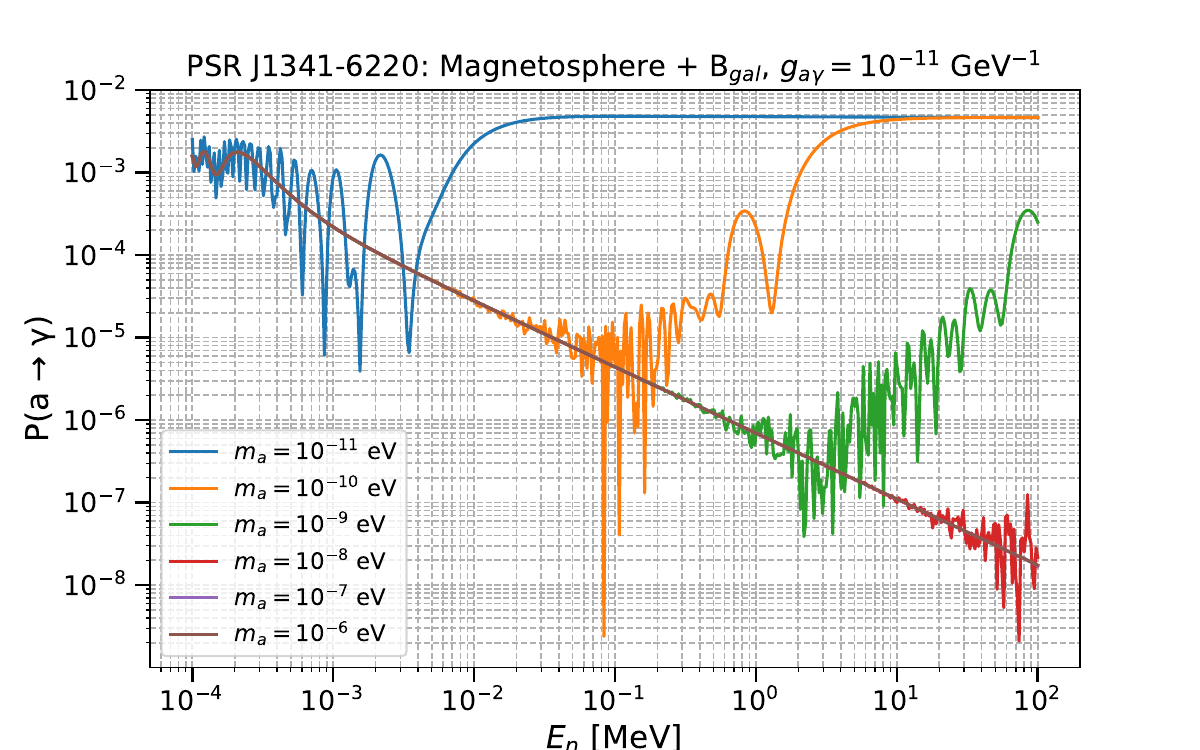}
\includegraphics[width=0.495\textwidth]{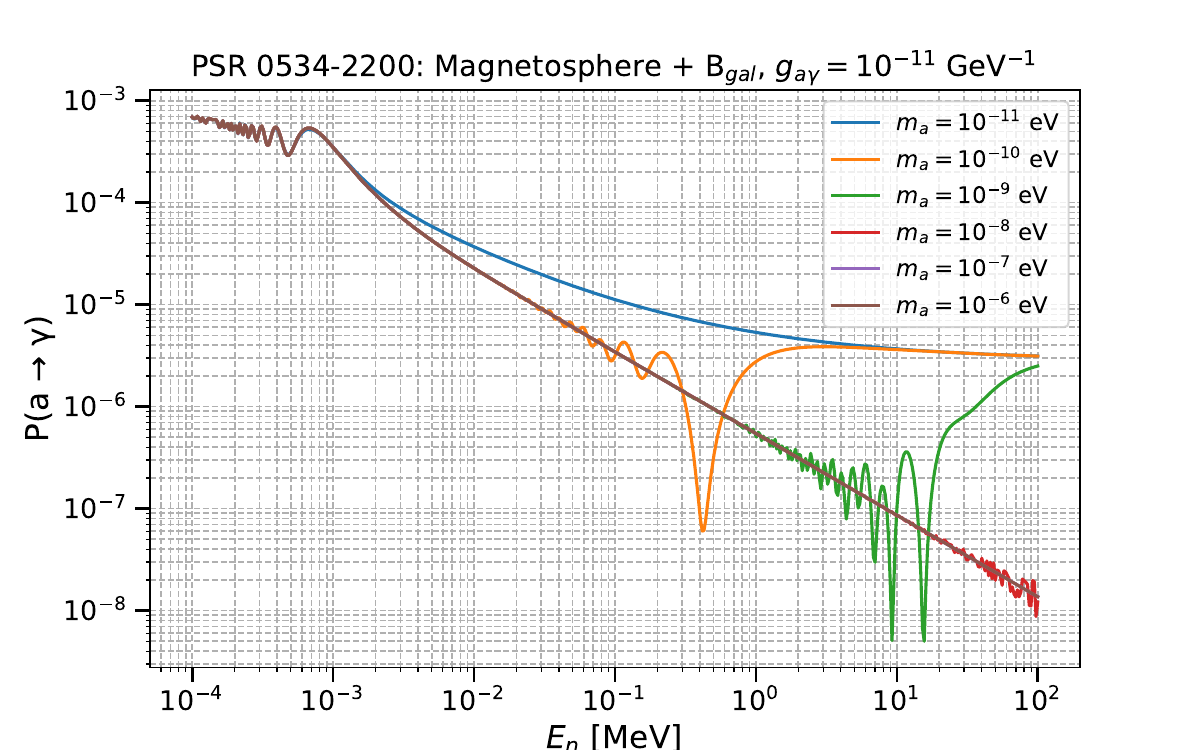}
\includegraphics[width=0.495\textwidth]{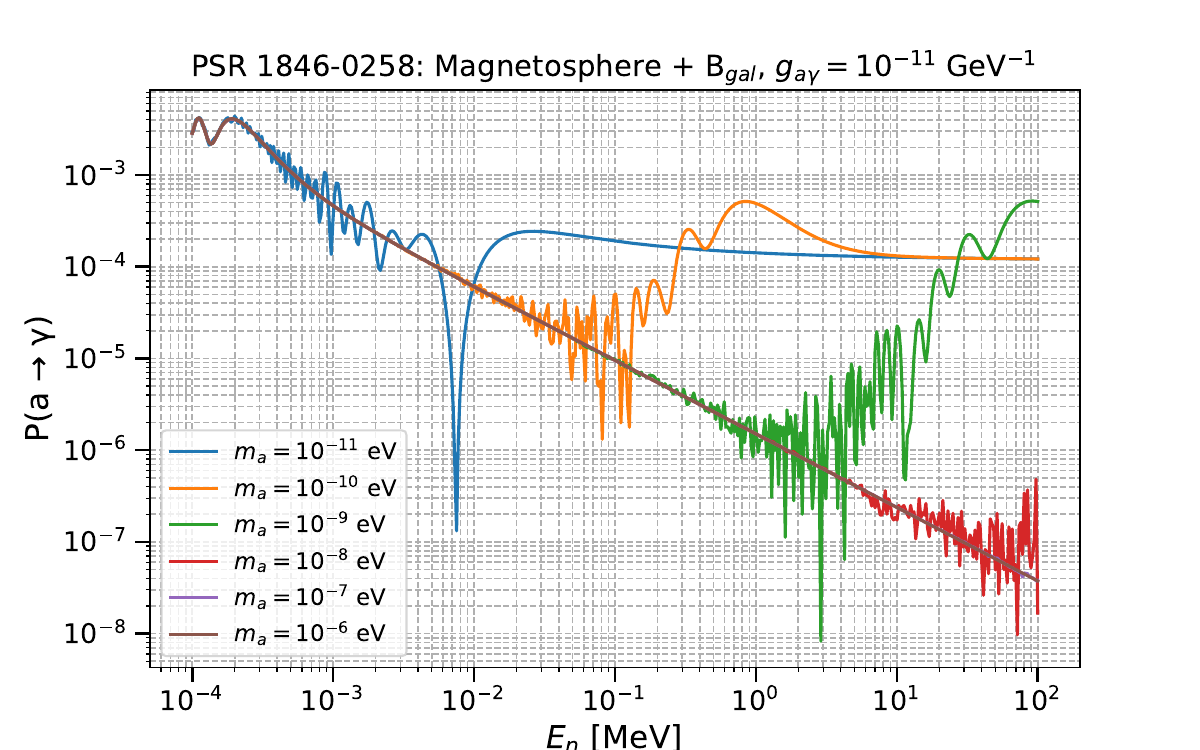}
\caption{Probability of conversion ALP$\rightarrow\gamma$ from ALP produced in PSR J1119-6127 (top left), PSR J1341-6220 (top right), PSR J0534-2200 (bottom left), PSR J1846-0258 (bottom right), for different masses from $m_a=10^{-11}$ eV to $m_a=10^{-6}$ eV, and fixed ALP-photon coupling $g_{a\gamma}=10^{-11}$ GeV$^{-1}$.}
\label{fig:totalprobabilityPSRs}
\end{figure}

In the following plots in Fig.~\ref{fig:totalprobabilityPSRs}, the total ALP-photon
conversion probability of all pulsars is shown. It is visible that small ALP masses
$m_a\leq 10^{-9}$~eV can convert in the Galactic magnetic field while higher ALP masses
are driven by conversion in the magnetosphere (shown by the overall diagonal decrease
with energy). Basically, as discussed in Sec.~\ref{conversioninside}, the mismatch term $\Delta_{\rm osc}$ is dominated by the plasma and
mass terms ($\Delta_{\rm pl}, \Delta_a \propto 1/E_n$) at low energy, and by the QED contribution
$\Delta_{\rm QED} \propto E_n B^2$ at higher energy, since the latter grows linearly with photon energy
while the former decrease. This produces a near-linear decline common to all masses at
intermediate-to-high energy --- where $\Delta_{\rm osc} \simeq \Delta_{\rm QED}$ and the curves overlap,
since $\Delta_{\rm QED}$ is mass-independent --- until, moving toward lower energy, the mass term
$\Delta_a$ becomes comparable to $\Delta_{\rm QED}$ (Eq.~(\ref{eq:oscillation-length})) and each curve departs from the common
trend, with the departure energy increasing with $m_a$. For $m_a \lesssim 10^{-9}\,{\rm eV}$, the rise
seen at even higher energy marks instead the separate transition from weak to strong mixing in the
Galactic magnetic field (Sec.~\ref{conversiongal}). Since $\omega_{\rm pl,gal} \sim 10^{-12}\,{\rm eV}$, efficient mixing
in the ISM is only achieved for such light ALPs, and again it is the mass value that sets where this
second transition ($E_{\rm cr}$) begins. 
Heavier masses remain magnetosphere-dominated over the whole energy range shown.
For example, comparing the pulsars, J1341-6220 has a larger conversion probability than J1119-6127,
reaching its maximum at lower energy, because it is 4~kpc more distant and the ALPs have a longer distance to convert. This can be understood from Eq.~\eqref{probabilityconversion}: in
the weak-mixing regime relevant here ($g_{a\gamma}=10^{-11}\,\mathrm{GeV}^{-1}$), for a
homogeneous path of length $L$ it reduces to the familiar two-level form
\begin{equation}
P_{a\rightarrow \gamma} \simeq \left(\frac{\Delta_{a\gamma}}{\Delta_{\rm osc}}\right)^2
\sin^2\!\left(\frac{\Delta_{\rm osc}L}{2}\right),
\label{eq:conversion-probability-approx}
\end{equation}
which, before saturation, grows with the square of the path length $L$. A more distant
pulsar therefore accumulates more mixing phase along its line of sight through $B_{gal}$, reaching saturation -- and thus its peak probability -- at lower energy, and with a higher saturation value.

%In the following plots in Fig.~\ref{fig:totalprobabilityPSRs} the total ALP-photon conversion probability of all pulsars is shown. It is visible that small ALP masses $m_a\leq 10^{-9}$ eV can convert in the Galactic magnetic field while higher ALP mass are driven by conversion in the magnetosphere (showed with the overall diagonal decreasing with energy). For example, comparing the pulsars, J1341-6220 has a larger conversion probability than J1119-6127, reaching its maximum at lower energy. This is because it is $4$ kpc more distant and the ALPs have more space to convert. \bjk{Why physically does the distance matter here?} \bjk{It might be useful to say a few more sentences about the structure of the curves in Fig. 5. Why is there an overall trend of the probability decreasing with energy? This comes from the magnetosphere conversion, right? Then at very high energy the galactic conversion becomes important?}

\subsection{Photon Spectrum}
\label{section:photonspectrum}

\begin{figure}[h!]
    \centering
    % Prima immagine
\includegraphics[width=0.495\textwidth]{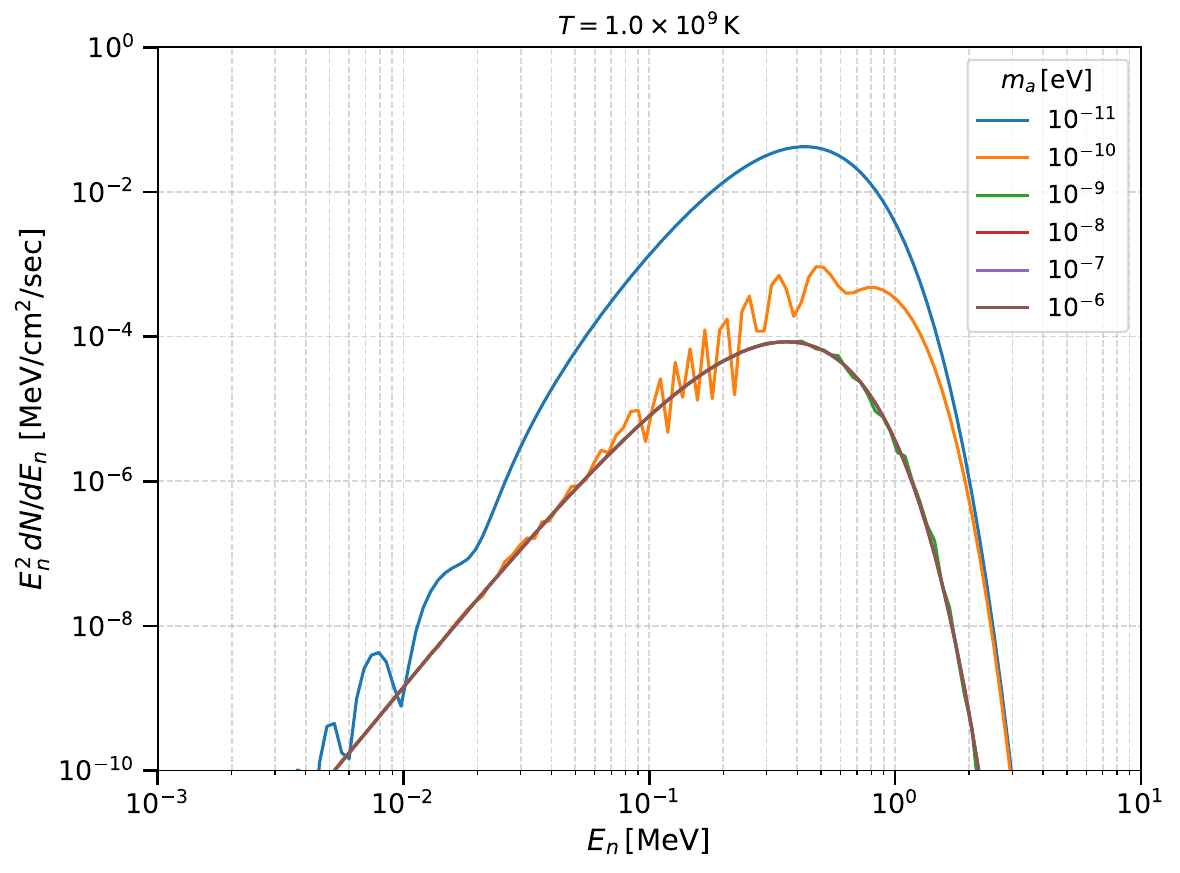}
\includegraphics[width=0.495\textwidth]{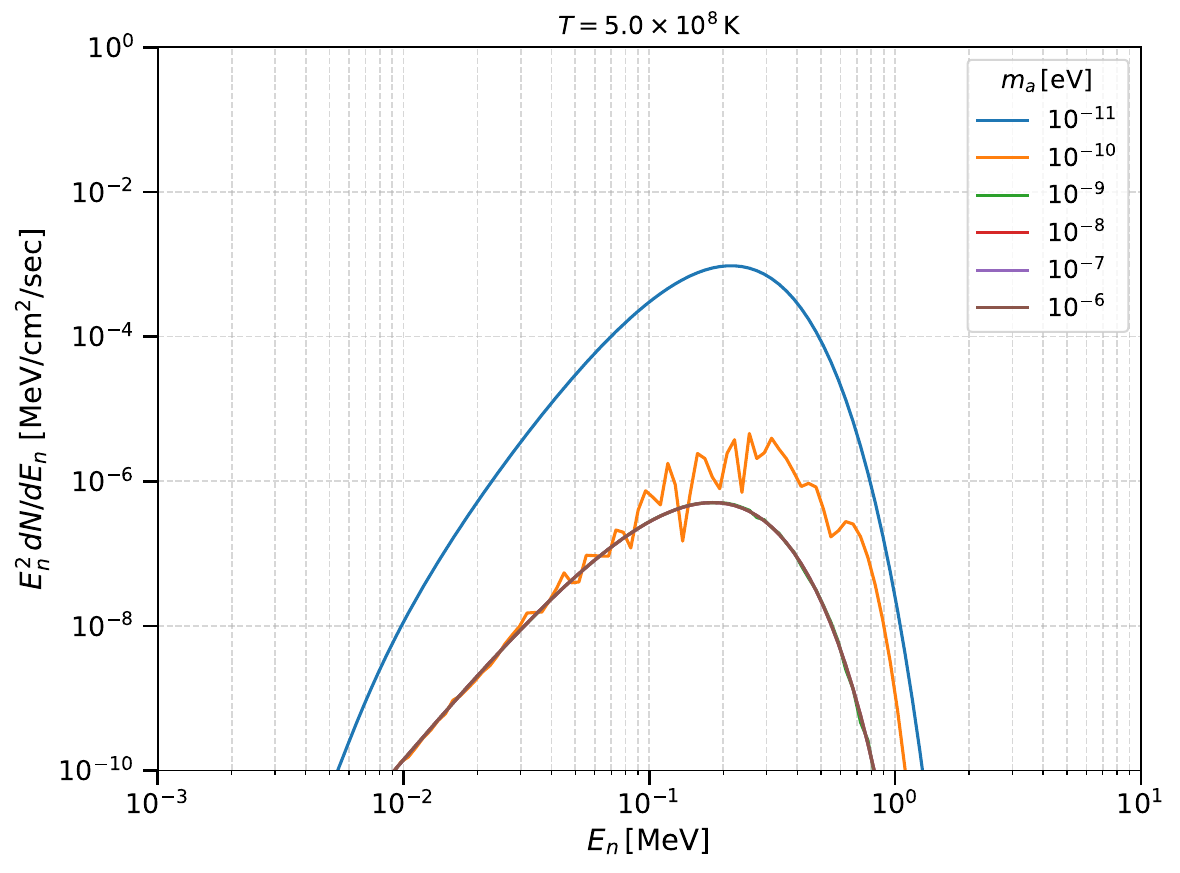}
\includegraphics[width=0.495\textwidth]{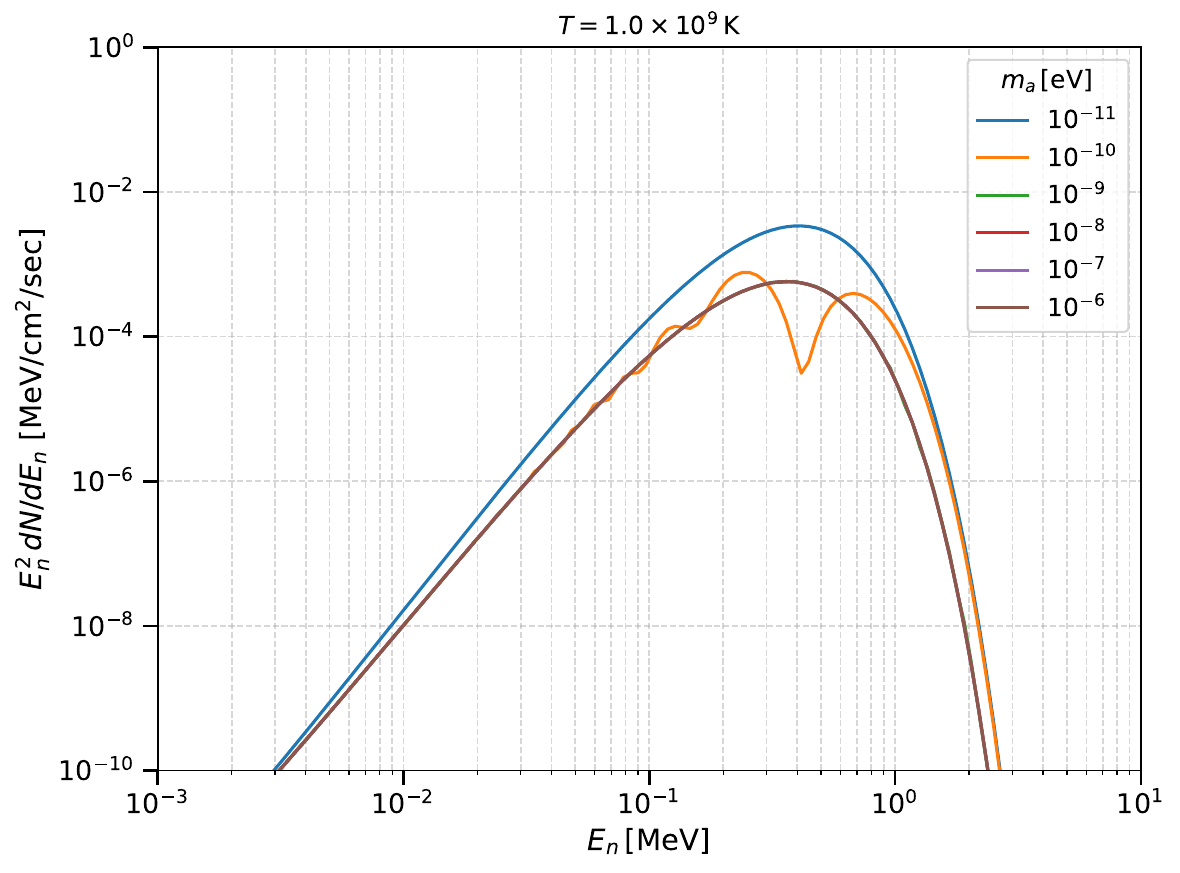}
\includegraphics[width=0.495\textwidth]{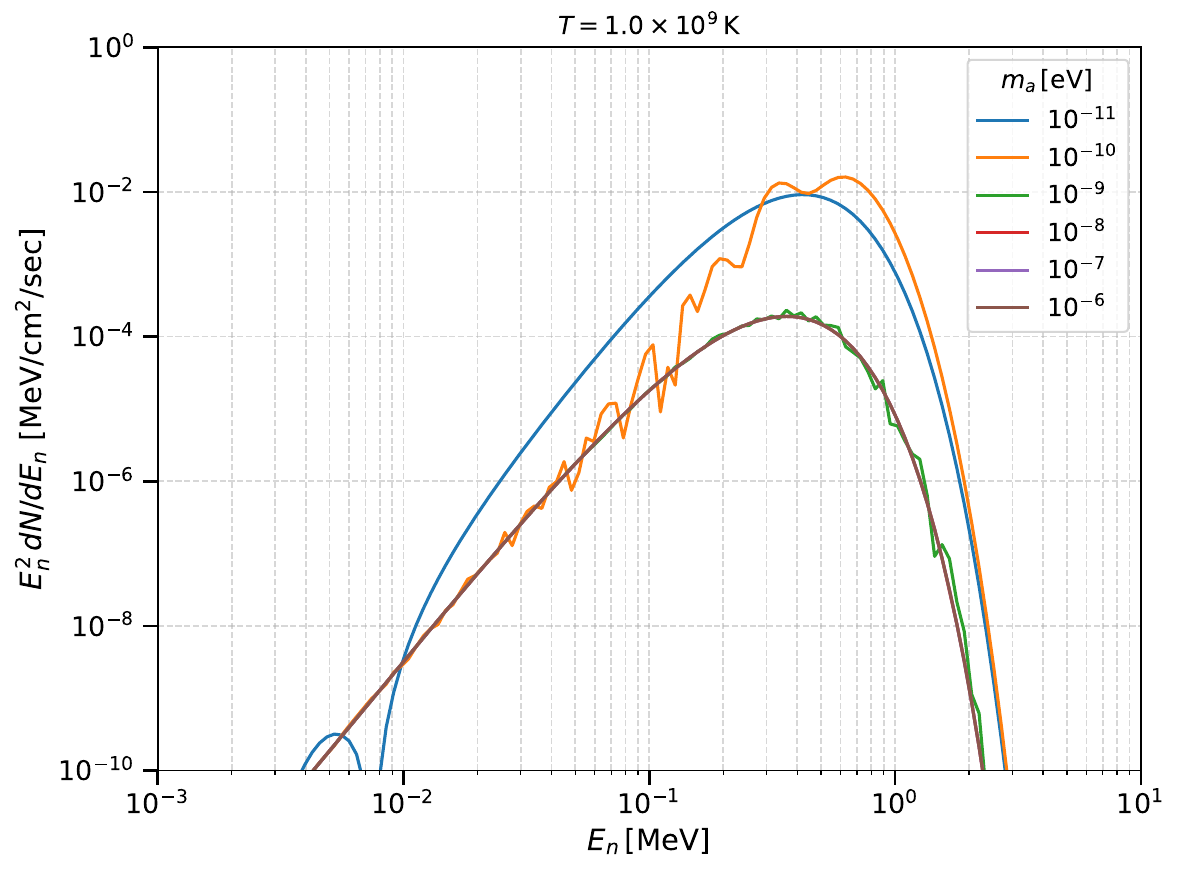}
\caption{\emph{Top left}: Observed flux of PSR J1119-6127 at $T=10^9$ K; \emph{Top right}: Observed flux of PSR J1341-6220 at $T=5\times10^8$ K; \emph{Bottom left}: Observed flux of PSR J0534-2200 at $T=10^9$ K; \emph{Bottom right}: Observed flux of PSR J1846-0258 at $T=5\times10^8$ K. All fluxes are computed for $g_{a\gamma}=10^{-11}$ GeV$^{-1}$ and six ALP mass values (from $10^{-11}$ to $10^{-6}$ eV).}
\label{fig:flux}
\end{figure}

The differential photon flux %(the number of the photon emitted in area, time and energy unit) 
is given by
\begin{equation}
   \frac{d\Phi_{\gamma}}{dE}=\frac{1}{4\pi d^2}\frac{dN_a}{dEdt}P_{a\rightarrow\gamma}(E)\,\,[\mathrm{MeV}^{-1}\,\mathrm{cm}^{-2}\,\mathrm{s}^{-1}]
\end{equation}
with $d$ the distance of the astrophysical object from the Earth. Using the ALP spectra produced in the NS interiors at the corresponding core temperatures of each PSR (given in Eq.~\eqref{temperature} and the surrounding text), and accounting for the total ALP-to-photon conversion probability, we obtain the resulting observable photon fluxes, shown in Fig.~\ref{fig:flux}.

\section{Inverse Compton (IC) and gamma-ray flux}
\label{section:IC}

In the previous section we have seen that the observed fluxes of gamma-rays from ALP-photon conversions (Fig.\ref{fig:flux}) do not reach the Fermi-LAT energies but cover regions of the so-called \emph{MeV gap experiments}. This band is not covered by any experiment. In fact, only NuSTAR operates at energies up to $80$ keV, not sufficient to explore these signals \cite{NuSTAR:2013yza, DeAngelis:2026wtq, Berge:2025kff}.

We consider ALPs thermally produced inside the neutron star at energies $E_a \lesssim 1$ MeV. These ALPs can subsequently convert into photons in the adiabatic regime ($E_a = E_\gamma$), as discussed in Section \ref{section:lpconversion}. The Inverse Compton process can upscatter MeV photons to energies detectable in the Fermi range. We now aim to determine whether a fraction of these photons undergoes
inverse Compton scattering in the pulsar magnetosphere and whether
the interaction occurs in the Thomson or Klein--Nishina regime
\cite{Blumenthal:1970gc,RybickiLightman1979}.

The pulsar magnetosphere is a region spatially defined (up to the light cylinder) by the $n_{GJ}$ charge density, as we have previously discussed. The electron number density is a crucial quantity because it determines the radius where ALP-photon conversion occurs; the number of electrons available for IC scattering; the optical depth in the magnetosphere; and whether the scattered photons can be boosted into the LAT energy range. The electron population consists of two components: primary electrons, which are either stripped from the stellar surface or created in the vacuum through electrostatic extraction due to the extremely high electric potential ($V_E \sim 10^{12} - 10^{16}$~V), and are accelerated within `gaps' (charge-starved regions such as the Polar Cap or the Outer Gap). These primaries emit ultra-high-energy photons via curvature radiation. These photons, by interacting with the strong magnetic field or other photons, produce $e^\pm$ pair cascades (the secondary component). According to the works  \citep{Daugherty:1995zy, Harding:2001gm, Hibschman:2001wk, Takata:2005tr, Hirotani:2007sq}, the energy distribution of the secondary component in the magnetosphere can be modeled as a broken power-law:
\begin{equation}
    n_e(\gamma) = N_0 \cdot \begin{cases}
\gamma^{-p_1} & \text{for} \quad\gamma_\mathrm{min} \le \gamma \le \gamma_\mathrm{break} \\
\gamma_\mathrm{break}^{p_2 - p_1} \gamma^{-p_2} & \text{for} \quad \gamma_\mathrm{break} < \gamma \le \gamma_\mathrm{max}
\end{cases}\,.
\end{equation}

The typical parameters for this distribution are: $p_1 \approx 1.5 \pm 0.5$ and $p_2 \approx 2.5 \pm 0.5$, representing the spectral indices; $\gamma_\mathrm{min} \approx 10 - 100$, related to the kinematic threshold for pair production; $\gamma_\mathrm{max} \approx 10^7 - 10^8$, limited by curvature radiation; and $\gamma_\mathrm{break} \approx 10^5$, associated with the equilibrium point between the cascade production rate and radiative losses. The normalization factor $N_0$ is linked to the pair multiplicity factor $\kappa$, such that the total numerical density is given by $n_e(r) = \kappa \,n_{GJ}(r)$. 

Inverse Compton scattering depends on both the photon energy $E_\gamma^i$ and the electron Lorentz factor $\gamma$.
For $\gamma E_\gamma^i \ll m_e c^2$, scattering occurs in the Thomson regime, where the average scattered photon energy is
\begin{equation}
E_\gamma^f \simeq \frac{4}{3}\gamma^2 E_\gamma^i,
\label{finalenergyIC}
\end{equation}
and the cross section is approximately constant, $\sigma_{\rm IC} \simeq \sigma_T= \frac{8 \pi \alpha^2}{3 m_e^2}$.

\begin{figure}[tb!]
    \centering
    \includegraphics[width=0.8\linewidth]{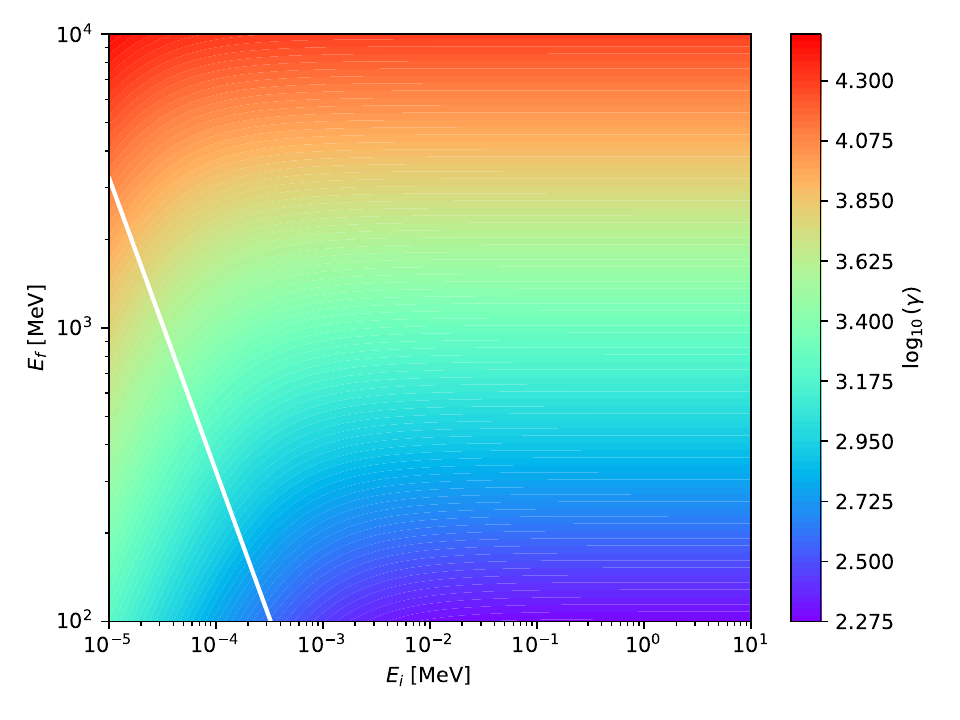}
    \caption{Contour plot of electron Lorentz factor $\gamma$ from Inverse Compton. The parameter space is given by initial photon energy produced in ALP-photon conversion $E^i_\gamma$ vs final photon energy in Fermi-LAT range $E^f_{\gamma}$. The color is the $\gamma$ given applying the Eq.~\eqref{finalenergyIC} if we are in Thomson regime, otherwise we have approximated with $\gamma=E_f/m_e$ in KN regime. The white line is the limit of the two regimes, $\gamma=m_e/E^i_{\gamma}$.}
    \label{fig:ICgamma}
\end{figure}

For $\gamma E_\gamma^i \gtrsim m_e c^2$, the interaction enters the Klein–Nishina regime, where the cross section is suppressed and the energy transfer saturates, with $E_\gamma^f \lesssim \gamma m_e c^2$. In this regime, the full Klein–Nishina differential cross section must be used. Thus, the condition to remain in the Thomson regime is $ \gamma < \frac{m_e}{E_\gamma^i}$ (white line in Fig.~\ref{fig:ICgamma}). This relation is particularly important because it must be evaluated at the conversion radius to determine whether the scattered photons can reach the Fermi-LAT energy band. 

To better visualize the regimes, we show in Fig.~\ref{fig:ICgamma} the contour plot of Eq.~\eqref{finalenergyIC} and the white line is the limit for the two regimes. Only the area below the white line is in Thomson regime and so only the initial photon energies in x-axis can reach be IC boosted at final photon energies in y-axis. 

From Fig.~\ref{fig:totalprod}, we estimate that the energy range to consider here is $E_\gamma^i \in [10^{-4},10]$ MeV and we will see that both regimes are relevant. In particular, keV photons can be efficiently up-scattered in the Thomson regime by electrons with Lorentz factors $\gamma \sim 10^2$, reaching the Fermi-LAT energy band ($\sim 100$ MeV). Higher-energy photons instead scatter in the Klein–Nishina (KN) regime, providing a subdominant contribution due to the reduced cross section. The inverse Compton emissivity is computed as
\begin{equation}
j(E_{\gamma}^f,r)=\frac{dN_{tot}}{dt\, dE_\gamma^fdV}(E_\gamma^f,r) =
\int d\gamma\, n_e(\gamma,r)
\int dE_\gamma^i \, n_\gamma(E_\gamma^i,r)
\frac{d\sigma_{\rm IC}}{dE_\gamma^f}(E_\gamma^i,E_\gamma^f,\gamma)\,,
\end{equation}
where we adopt the full Klein–Nishina differential cross section following \cite{Blumenthal:1970gc}
\begin{equation}
    \frac{d\sigma_{IC}}{dE^f_{\gamma}}=\frac{3\sigma_T}{4\gamma^2 E^i_{\gamma}} F(q,\Gamma)\,,
\end{equation}
where $q=E^f_{\gamma}/(\Gamma(\gamma m_e-E^f_{\gamma}))$ and $\Gamma=4\gamma E^i_{\gamma}/m_e$, defining the function $F(q,\Gamma)$ as 
\begin{equation}
F(q,\Gamma)=2q\ln (q)+(1+2q)(1-q)+\frac{(\Gamma q)^2(1-q)}{2(1+\Gamma q)}, \quad 0<q\leq1 .
\end{equation}
This formalism naturally includes both the Thomson and Klein–Nishina regimes without requiring an a priori separation. The limits on $q$ have a clear physical and kinematical origin and imply constraints on the final photon energy $E_f$.

The target photon density $n_\gamma(E_\gamma^i,r)$ is related to the photon injection rate from ALP conversion through
\begin{equation}
\frac{d N_\gamma}{dEdt} =
P_{a\to\gamma}(E)\, \frac{dN_a}{dE\, dt},
\end{equation}
and, assuming radial propagation, can be expressed as
\begin{equation}
n_\gamma(E,r) \simeq \frac{1}{4\pi r^2}\, \frac{d N_\gamma}{dEdt}.
\end{equation}
Finally, the observable differential photon flux at Earth is
\begin{equation}
\frac{d\Phi_\gamma(E_\gamma^f) }{dt dE_{\gamma}^f}=
\frac{1}{4\pi d^2}
\int^{R_\mathrm{max}=10^4r_0}_{r_0} d^3r\,
\frac{dN_{tot}}{dt\, dE_\gamma^f}(E_\gamma^f,r),
\label{eq:ICflux}
\end{equation}
which can be directly compared with Fermi-LAT measurements.

\begin{figure}
    \centering
    \includegraphics[width=0.7\linewidth]{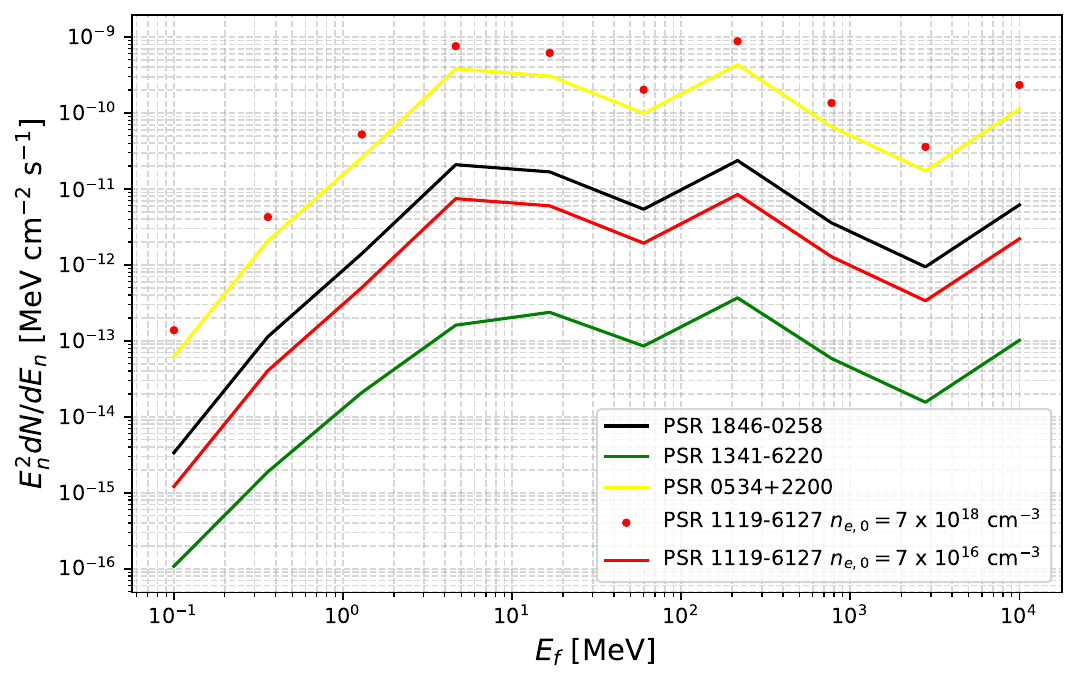}
    \caption{Photon flux after Inverse Compton for the four PSRs applying Eq.~\eqref{eq:ICflux}, for $m_a=10^{-11}$ eV and $g_{a\gamma}=10^{-11}$ GeV$^{-1}$. Only for PSR 1119-6127, the photon flux is shown for two different values of $n_{e,0}$, ($7\times 10^{16}$ cm$^{-3}$, $7\times 10^{18}$ cm$^{-3}$).}
    \label{fig:ICFLUX}
\end{figure}

We see in Fig.~\ref{fig:ICFLUX} that the IC boosts the photon flux at energies $>100$ MeV but the intensity is at the order of $10^{-11}$ MeV/cm$^2$/s, meaning that the ALP contribution is negligible. The total optical depth, which quantifies the dimming of the source, is $3.5\times10^{-3}$ throughout the magnetosphere, where the maximum is near the surface.

%For PSR J1119-6127, we compute the photon injection rate for different values of $r_{\rm max}$ ($10^2 r_0$, $10^3 r_0$, $10^4 r_0$), ALP masses ($10^{-6}$ eV and $10^{-11}$ eV), neutron star temperatures ($10^8$–$10^9$ K), and couplings ($10^{-11}$–$10^{-13}$ GeV$^{-1}$), obtaining values in the range $1.4\times 10^{35}$ s$^{-1}$ to $7.3\times 10^{36}$ s$^{-1}$.

\subsection{Absorption of photons}
\label{section:absorption}
The propagation of photons within the pulsar magnetosphere is governed by energy-dependent absorption regimes that necessitate a clear distinction between pre- and post-inverse Compton (IC) scattering stages. 
For low-energy photons ($E_\gamma \lesssim 1$ MeV), the primary attenuation mechanism is resonant cyclotron absorption, which occurs when the photon frequency matches the local Larmor frequency of magnetospheric electrons. Conversely, photons up scattered to high energies ($1 \text{ MeV} < E_\gamma < 10^4 \text{ MeV}$) via IC processes enter a kinematic regime where magnetic pair production ($\gamma + B \to e^+ e^-$) and photon-photon annihilation ($\gamma \gamma \to e^+ e^-$) dominate the opacity. %Since the IC boost fundamentally redistributes the photon population across these thresholds, we separately parametrize the survival probability using pre-IC factors for the primary flux and post-IC attenuation for the boosted component to accurately determine the final observable spectrum.

\newpage
%\subsubsection{Absorption pre-IC}

After ALP--photon conversion in the magnetosphere, the photon flux may be attenuated before undergoing inverse Compton  scattering. We parametrize this effect through a pre-IC survival factor
\begin{equation}
G_{\rm pre}^{\rm cyc}(E_\gamma)=\exp[-\tau_{\rm cyc}(E_\gamma)].
\end{equation}
%For non-relativistic magnetospheric electrons, the resonant cyclotron absorption coefficient can be written as \[\tau_{\rm pre}(E)=\int_{r_{\rm conv}}^{r_{\rm IC}}\left[\alpha_{\rm cyc}(E,r)+n_e(r)\sigma_{\rm C}(E)+\alpha_{\rm ff}(E,r)\right]dr .\]
In the region relevant for our model, $(r_{\rm conv}\sim 10^3-10^4 r_0)$, the plasma density and magnetic field have already decreased substantially, so ordinary compton scattering and free--free absorption are expected to be small. The main possible suppression before IC comes from resonant cyclotron absorption
\begin{equation}
E_\gamma \simeq E_B=\hbar\omega_B(r)=\hbar\frac{eB(r)}{m_ec}.
\end{equation}
So only if the distance $r_\mathrm{res}=r_0(B_0/(m_eE_{\gamma}))^{1/3}$ (when $E_\gamma \simeq E_B$) is between $r_\mathrm{conv}$ and $r_{IC}$,
\begin{equation}
\tau_{\rm cyc}(E_\gamma)=
\int_{r_{\rm conv}}^{r_{\rm IC}}
\alpha_{\rm cyc}(E_\gamma,r)\,dr = \int_{r_{\rm conv}}^{r_{\rm IC}} \frac{\pi^2}{m^2_e} n_e(r)(1+\cos^2\theta)\delta(E_{\gamma}-E_B).
\end{equation}
If \(G_{\rm pre}\simeq 1\) or $\tau\ll1$, the magnetosphere is effectively transparent to the photons, and if \(G_{\rm pre}<1\) or $\tau\gg 1$ a cyclotron resonance is crossed and the photon is absorbed.

Looking at Fig.~\ref{ratio} for all PSRs, we see that only the Crab for energy $0.1$ keV has a $r_\mathrm{conv}\simeq r_\mathrm{res}$. For example, for PSR 1116-6127 at $E_{\gamma}=1$ keV the conversion radius is $5\times 10^8$~cm and at $0.5$ MeV it is $r_\mathrm{conv}=10^{10}$~cm. In order to have the absorption at the same energies we require that $r_\mathrm{res}>r_\mathrm{conv}$ but this is not the case: $r_\mathrm{res}=1.2\times10^7$~cm and $r_\mathrm{res}=1.5\times 10^6$~cm, respectively. We therefore do not expect the absorption of photons from ALP conversion in these systems (with only the Crab being at the limit of possible absorption) and we thus neglect this effect in the following.

%\subsubsection{Absorption post-IC}

\section{Results}
\label{section:results}

Comparing the Fermi-LAT data and the photon flux from ALP conversion, the ALP parameter space cannot be tested by Fermi-LAT. However, the no-IC flux can be tested by MeV experiments.  From the analysis of COMPTEL data, we find that only the Crab pulsar has been observed in the MeV energy range.  For the COMPTEL datasets we test the ALP-photon
conversion hypothesis against the intrinsic pulsar spectrum using a
profile-likelihood ratio test.

For the COMPTEL dataset (Table 3 of \cite{Kuiper:2001ev}), the observed differential photon flux $F_{\rm obs}(E)$ is compared directly to the model
\begin{equation}
    F_{\rm tot}(E) = F_{\rm astro}(E; N_0) + F_{\rm ALP}(E; g_{a\gamma}, m_a),
\end{equation}
where $F_{\rm ALP}$ is the photon flux from ALP-photon conversion and $F_{\rm astro}$ is a broken power law with fixed break energy and spectral indices (\cite{Kuiper:2001ev}) and free normalization $N_0$.

%For the INTEGRAL dataset (JEM-X, ISGRI, SPI, \cite{Mineo:2006jh}), the measured quantities are background-subtracted instrumental count rates ($\mathrm{ph\,s^{-1}\,keV^{-1}}$) not physical fluxes. We therefore forward-fold the predicted physical flux through the energy-dependent effective area $A_{\rm eff}^{(i)}(E)$ of each instrument $i \in \{\mathrm{JEM\text{-}X, ISGRI, SPI}\}$,
%\begin{equation}N_{\rm pred}^{(i)}(E) = \left[F_{\rm astro}(E; K, a) + F_{\rm ALP}(E; g_{a\gamma}, m_a)\right]\times A_{\rm eff}^{(i)}(E),
%\end{equation}

%where the intrinsic spectrum follows the log-parabolic form of \cite{Mineo:2006jh}, $F_{\rm astro}(E) = K\,E^{-(a + b\log_{10}E)}$, with $K$ and $a$ free and the curvature $b=0.14$. 

\begin{figure}[tb!]
    \centering
    \includegraphics[width=1\linewidth]{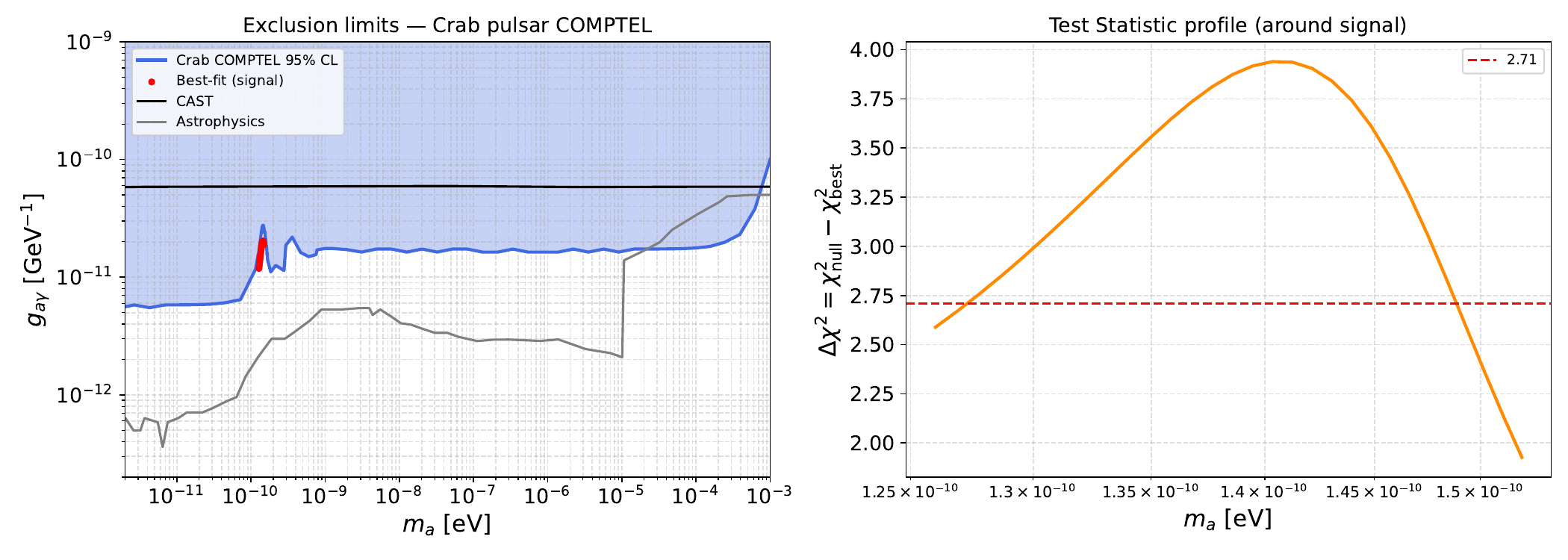}
    \caption{Constraints on the ALP-photon coupling derived from the COMPTEL observations of the Crab pulsar \cite{Kuiper:2001ev}. \emph{Left panel}: 95\% confidence level upper limits on the ALP-photon coupling $g^{UL}_{a\gamma}$ as a function of the ALP mass $m_a$, obtained by fitting the COMPTEL spectrum with a broken power-law \cite{Kuiper:2001ev} plus an ALP-induced component. The blue shaded region is excluded by this analysis. Red markers indicate the best-fit values $g^{best}_{a\gamma}$ for masses with $\mathrm{TS}>2.71$. We compare the result with CAST (black) \cite{CAST:2024eil} and all astrophysical bounds (gray) in \cite{Fermi-LAT:2016nkz, Noordhuis:2022ljw, Li:2024zst, Ning:2024eky}. \emph{Right pane}l: Test Statistic profile as a function of the ALP mass of the red points. The horizontal dashed line corresponds to $\mathrm{TS}=2.71$, which is adopted as the reference threshold for a one-sided $95\%$ confidence level excess according to Wilks' theorem.}
    \label{fig:comptelCrab}
\end{figure}

For each fixed ALP mass $m_a$ and coupling $g_{a\gamma}$, the nuisance
parameters ($N_0$ for COMPTEL%; K, a$ for INTEGRAL
) 
are profiled by
minimizing the $\chi^2$ (equivalently, $-2\ln\mathcal{L}$ for Gaussian
uncertainties) using the MIGRAD algorithm~\citep{James:1975dr}, as implemented in the
\texttt{iminuit} Python package~\citep{iminuit}. The test statistic at each mass is defined as
\begin{equation}
    TS(m_a) = \chi^2_{\rm null} - \chi^2_{\rm best}(m_a),
\end{equation}
where $\chi^2_{\rm null}$ is the null-hypothesis fit ($g_{a\gamma}=0$)
and $\chi^2_{\rm best}(m_a) = \min_{g_{a\gamma}} \chi^2(g_{a\gamma}, m_a)$.
The coupling value minimizing $\chi^2$ at fixed mass is denoted
$g_{a\gamma}^{\rm best}(m_a)$. The 95\% CL upper limit,
$g_{a\gamma}^{\rm UL}(m_a)$, is defined as the value of $g_{a\gamma}$
satisfying

\begin{equation}
    \chi^2\!\left(g_{a\gamma}^{\rm UL}\right) = \chi^2_{\rm best} + 2.71,
\end{equation}
the asymptotic one-sided 95\% CL threshold for one degree of freedom
with the physical parameter bounded at $g_{a\gamma}\geq0$~\cite{Chernoff:1954eli}. This threshold is located directly on the $\chi^2(g_{a\gamma})$
curve, evaluated on a dense logarithmic grid in $g_{a\gamma}$, rather
than through a local numerical search, to avoid instabilities from the
non-smooth (oscillatory) behaviour of $P_{a\to\gamma}$.
We stress that the best-fit points shown in Fig.~\ref{fig:comptelCrab} do not constitute evidence for an ALP contribution to the Crab spectrum: given the simplicity of the adopted broken-power-law model for the intrinsic pulsar emission, a marginal improvement in $\chi^2$ at $TS>2.71$ is expected to arise from spectral features not captured by this simplified astrophysical model, rather than from genuine ALP-induced emission. We therefore do not claim a detection, and these red points should be interpreted purely as the outcome of the profile-likelihood procedure at masses where the fit statistically favours $g_{a\gamma}>0$. Furthermore, $TS(m_a)$ exceeds the $2.71$ threshold only within a narrow mass window around $m_a\sim10^{-10}\,{\rm eV}$ (Fig.~\ref{fig:comptelCrab}, right panel), which is why best-fit points are reported only in this range. This behaviour follows directly from the oscillatory, resonance-driven structure of $P_{a\to\gamma}(E;m_a,g_{a\gamma})$ discussed in Sec.~\ref{section:lpconversion}: outside this narrow mass range the predicted ALP spectral feature no longer overlaps in energy with the COMPTEL data points in a way that improves the fit, so $TS(m_a)$ drops below threshold and only an upper limit — not a best-fit value — can be meaningfully defined. In most of the mass range probed by COMPTEL, the new limits are stronger than the CAST bound but weaker than existing astrophysical constraints. For $m_a \gtrsim 10^{-5}\,{\rm eV}$, however, the COMPTEL upper limits begin to exclude previously unconstrained parameter space.

\begin{figure}[tb!]
    \centering
\includegraphics[width=0.495\textwidth]{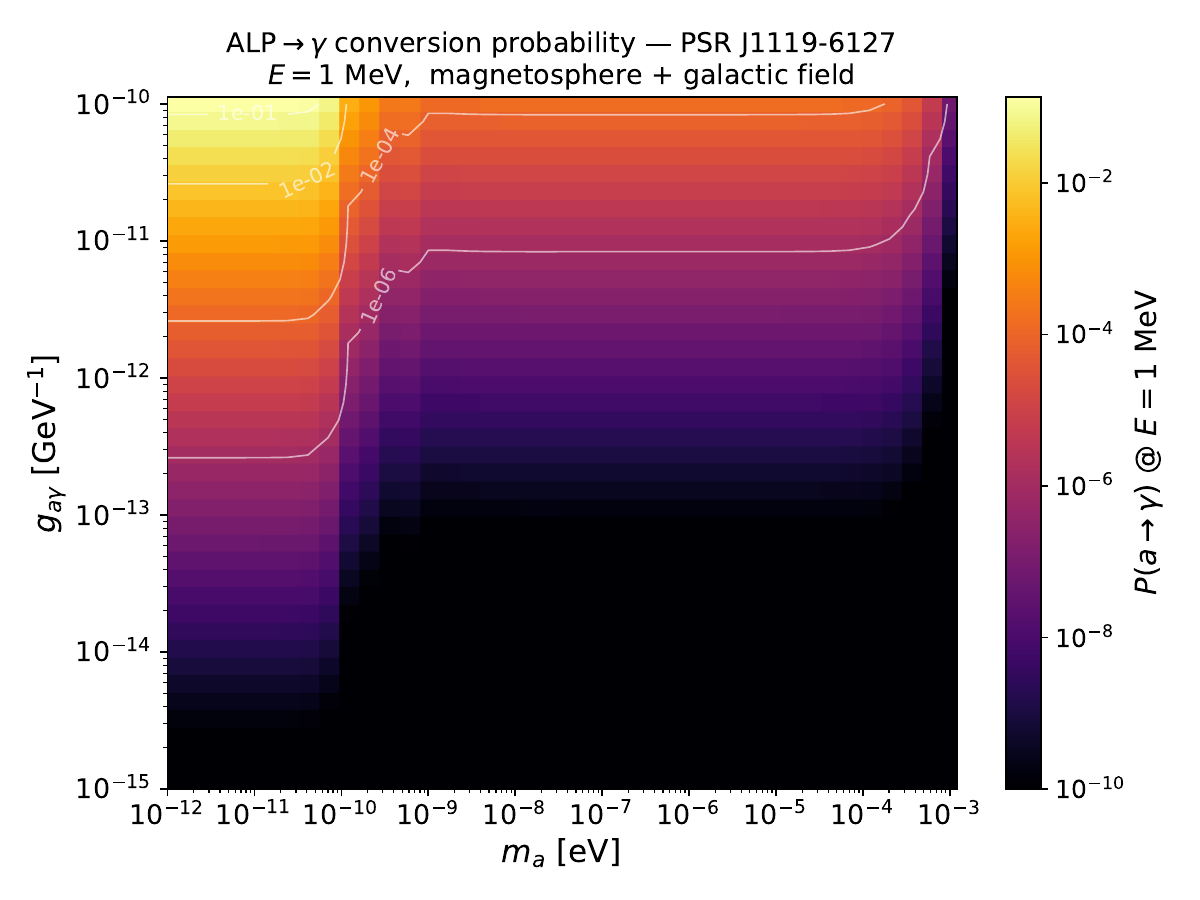}
\includegraphics[width=0.495\textwidth]{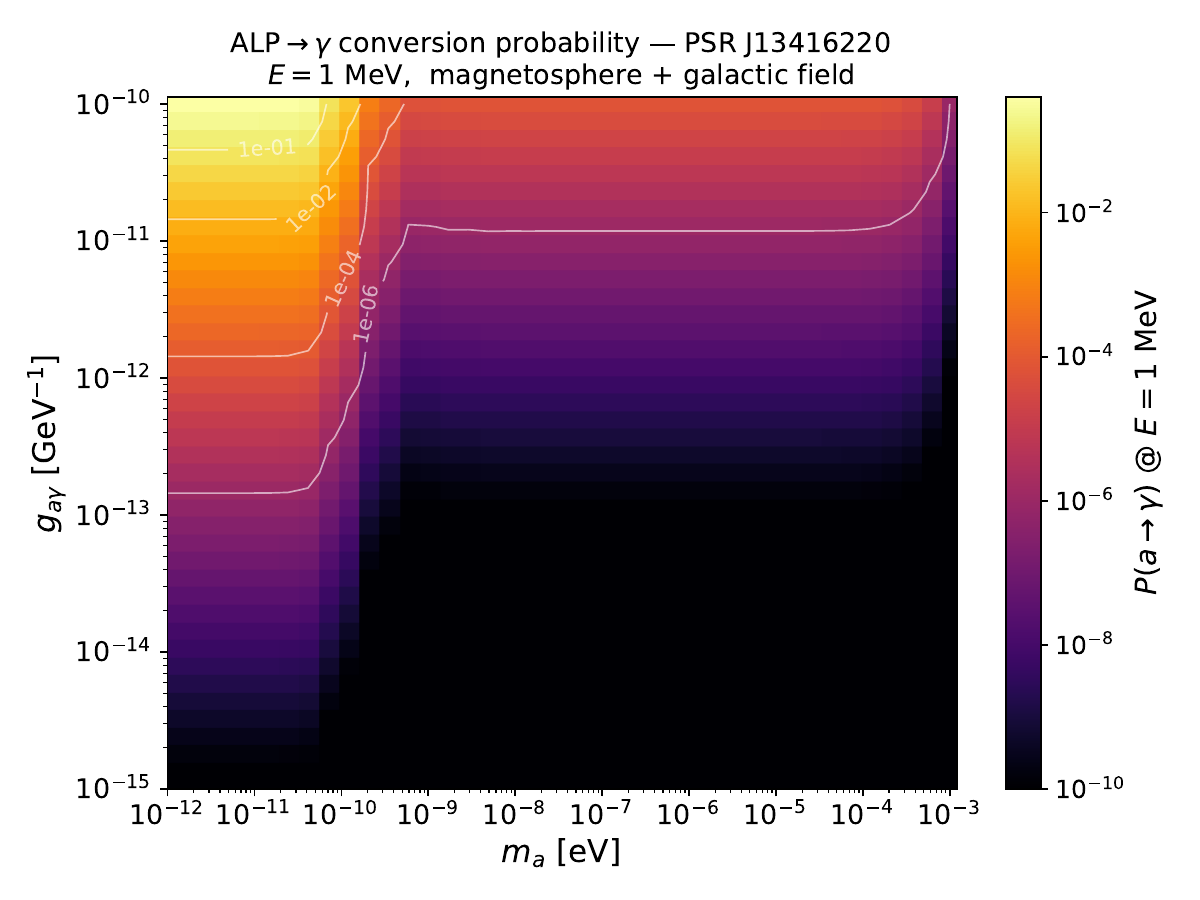}
\includegraphics[width=0.495\textwidth]{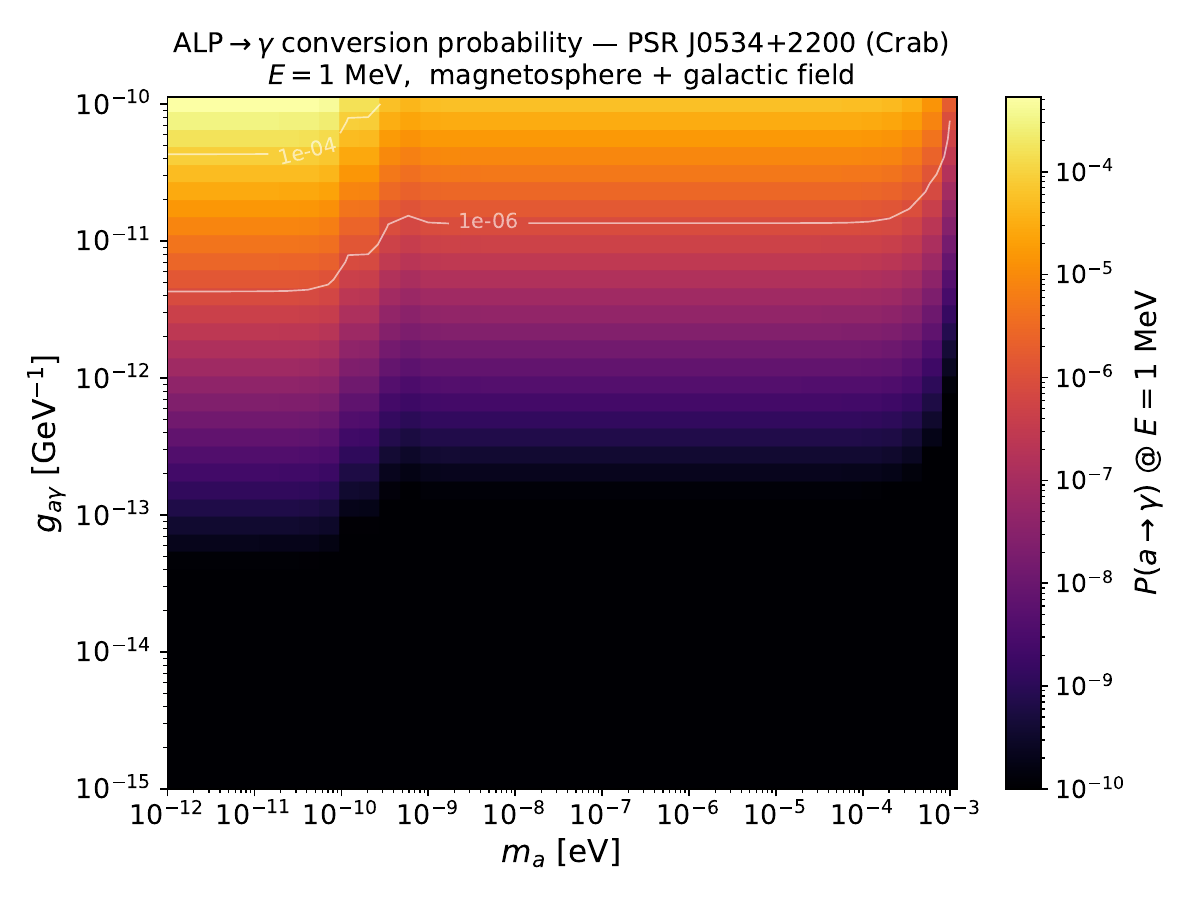}
\includegraphics[width=0.495\textwidth]{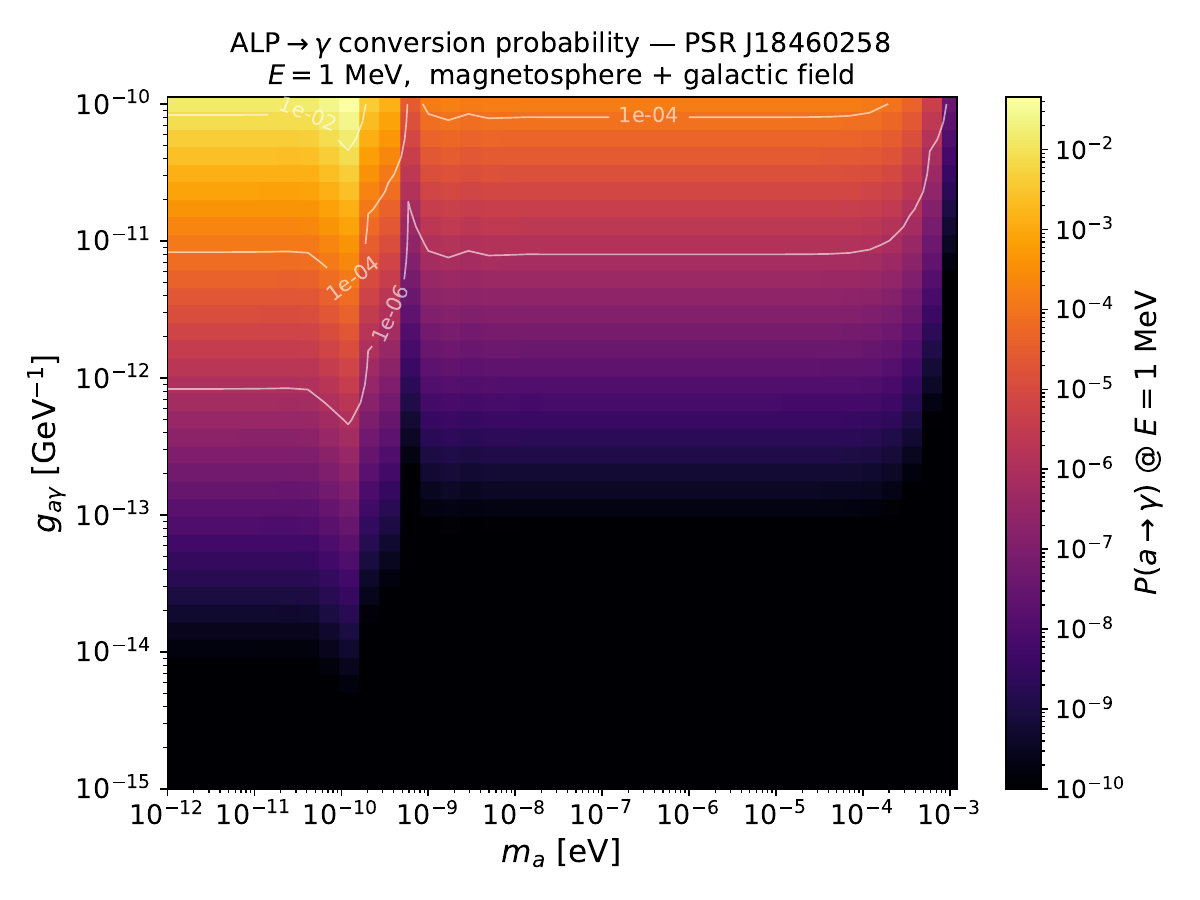}
\caption{Photon conversion probability $P(E;a\rightarrow\gamma)$ for energy $E=1$ MeV, as a function of the ALP mass $m_a$ [eV] and the photon--ALP coupling $g_{a\gamma}$ [GeV$^{-1}$].}
    \label{fig:probability1MeV}
\end{figure}

\begin{figure}[htbp!]
    \centering
    \includegraphics[width=1\linewidth]{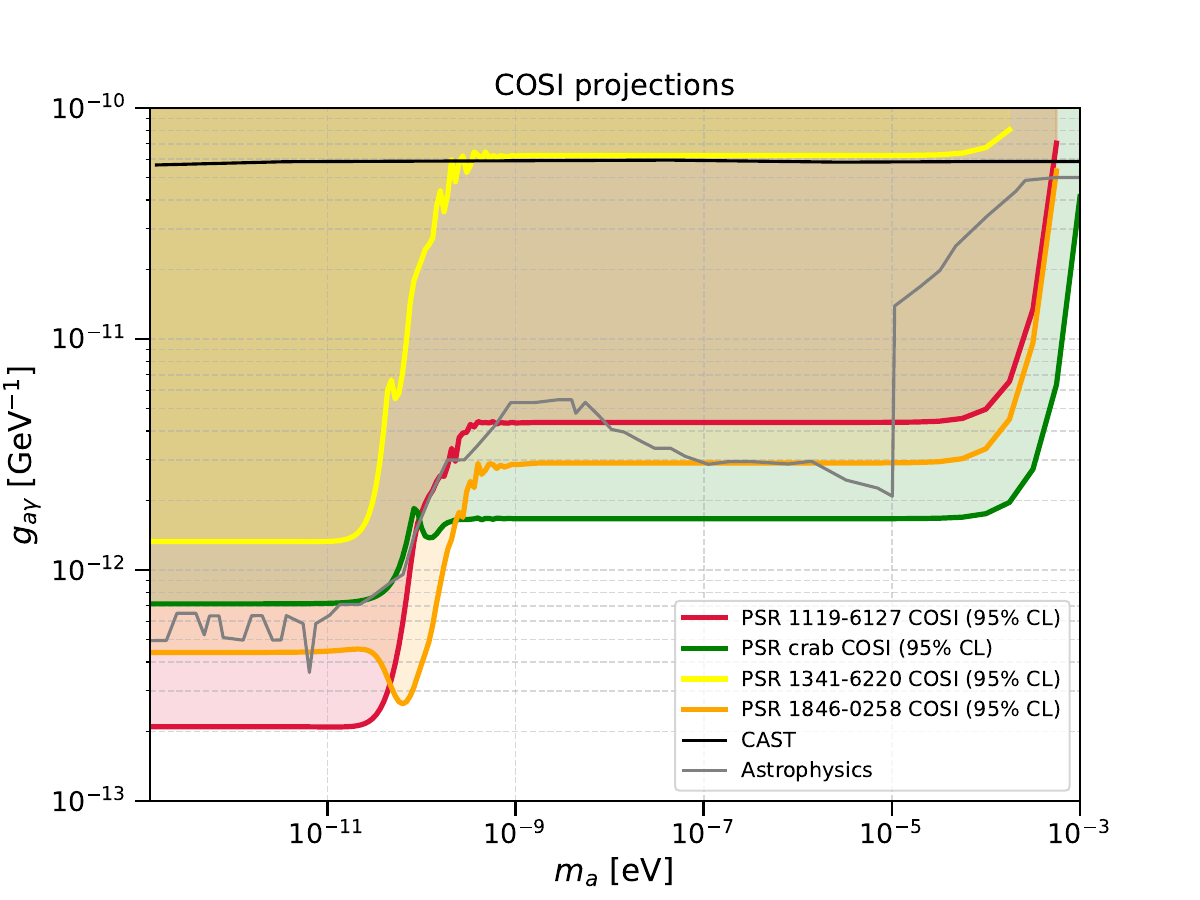}
    \caption{Parameter space tested at $95\%$ C.L. by COSI after $T_{obs}=2$ yr of PSR 1119-6127 (red), Crab (green), PSR 1341-6220 (yellow), PSR 1846-0258 (orange). CAST (black)\cite{CAST:2024eil}, all astrophysical bounds (gray) in \cite{Fermi-LAT:2016nkz, Noordhuis:2022ljw, Li:2024zst, Ning:2024eky}. The other couplings are fixed at $g_{ap}=g_{an}=5\times10^{-9}\text{ GeV}^{-1}$ and $g_{ae}=1.3\times10^{-13}\text{ GeV}^{-1}$.}
    \label{fig:placeholder}
\end{figure}

The no-IC flux component can also be probed by the future NASA Small Explorer mission, the Compton Spectrometer and Imager (COSI) \cite{Kierans:2022eid,Beechert:2022phz,COSI:2026asd}. Scheduled for launch into low-Earth orbit in 2027, COSI is a wide-field gamma-ray telescope designed to survey the entire sky in the (0.2-5) MeV energy range. In Fig.~\ref{fig:placeholder}, the projected $95\%$ C.L. sensitivity of COSI after two years of observations for PSR J1119-6127, the Crab pulsar, PSR J1846-0258, and PSR J1341-6220 was derived through a direct profile-likelihood scan over the ALP parameter space ($m_a$, $g_{a\gamma}$). The analysis is based on an Asimov dataset~\cite{Cowan:2010js}, in which the observed counts are assumed to be equal to the expected background. For each ALP mass, the likelihood was evaluated over a grid of coupling values, and the projected upper limit was determined by identifying the value of $g_{a\gamma}$ for which $\Delta(-\ln\mathcal{L}) = 1.35$, corresponding to the one-sided $95\%$ confidence interval. This approach avoids the numerical instabilities that may arise in minimization-based methods while providing a robust determination of the sensitivity. The resulting limits are shown in Fig.~\ref{fig:placeholder}, together with the current CAST constraints \cite{CAST:2024eil} and the existing astrophysical bounds reported in \cite{Fermi-LAT:2016nkz, Noordhuis:2022ljw, Li:2024zst, Ning:2024eky}. The projected sensitivities are consistent with the ALP fluxes presented in Fig.~\ref{fig:flux} and with the behaviour of the photon conversion probability shown in Fig.\ref{fig:probability1MeV} for $E=1$ MeV. For low ALP masses ($m_a\simeq 10^{-11}$ eV), the predicted fluxes are influenced by the conversion in the galactic magnetic field. At larger masses, the ALP-photon conversion probability is driven by the magnetosphere. The Crab pulsar, being the closest source in our sample, produces the highest observable ALP flux, allowing COSI to probe smaller values of $g_{a\gamma}$, whereas more distant pulsars yield progressively weaker sensitivities.

%%%%%%%%%%%%%%%%%%%%%%%%%%%%%%%%%%%%%%%%%%
\section{Conclusions}
\label{section:conclusion}

In this work, we have investigated the possibility of probing ALPs through their production in neutron stars and subsequent conversion into photons in neutron-star magnetospheres and in the Galactic magnetic field. We have focused on four representative pulsars spanning different astrophysical regimes, including the Crab pulsar, the magnetar-like pulsar PSR J1119–6127, PSR J1846–0258, and the older and more distant PSR J1341–6220. Our analysis provides a unified framework connecting ALP production in the dense stellar interior with photon conversion and subsequent radiative reprocessing in the magnetospheric environment.

The ALP production calculations show that, for the temperatures relevant to young neutron stars, nucleon–nucleon bremsstrahlung provides the dominant contribution to the ALP emissivity over the other production channels considered in this work. In particular, the characteristic core temperatures of young pulsars and magnetar-like objects are expected to be of order $10^9$ K, implying that their internal ALP production can be comparable even when their surface magnetic and X-ray properties are substantially different. The resulting ALP spectra are predominantly concentrated at energies below the MeV scale, while the thermal suppression of the production rate strongly limits the intrinsic flux at energies approaching the Fermi-LAT range.

We have shown that the subsequent ALP–photon conversion is controlled by the interplay between the ALP mass, the plasma density, and QED vacuum birefringence. In the inner magnetosphere, the strong magnetic field does not necessarily maximize the conversion efficiency, since the QED contribution to the photon dispersion relation can strongly suppress ALP–photon mixing close to the stellar surface. Efficient conversion instead develops at larger distances, where the dipolar magnetic field has decreased sufficiently for the QED suppression to become weaker. For sufficiently light ALPs, with masses around or below $10^{-9}$ eV, conversion in the Galactic magnetic field can provide an additional and potentially important contribution to the observable signal. The relative importance of magnetospheric and Galactic conversion therefore depends strongly on the ALP mass and on the distance to the source.

The photons produced by ALP conversion naturally populate the hard X-ray and MeV energy range. We have further investigated whether inverse Compton scattering in the pulsar magnetosphere can shift part of this emission to higher energies. Although the inverse Compton process can in principle boost the photon energies above $100$ MeV, our results indicate that the resulting ALP-induced flux remains too small to provide a detectable contribution in the Fermi-LAT band for the benchmark scenarios considered here. Consequently, current Fermi-LAT observations do not provide significant sensitivity to the ALP parameter space explored in this study. This result highlights the importance of the largely unexplored MeV energy range, where the primary photons produced through ALP conversion are expected to be more relevant.

Existing hard X-ray and MeV observations, including archival COMPTEL data, can already constrain part of the ALP parameter space, although the sensitivity is limited by the instrumental coverage and by the astrophysical backgrounds. Our projections show that future MeV observations, in particular with COSI, could substantially improve these constraints. The expected sensitivity is especially promising for light ALPs, for which conversion in the Galactic magnetic field can enhance the signal, while heavier ALPs are predominantly constrained through conversion in the neutron-star magnetosphere. The combination of different pulsar targets may further improve the robustness of the search, since the conversion probability depends on the magnetic-field strength, source distance, and magnetospheric environment.

Overall, our results demonstrate that young pulsars and magnetar-like neutron stars constitute promising laboratories for studying ALP–photon interactions, not because they necessarily produce a detectable signal in the GeV band, but because they provide a natural connection between efficient ALP production in dense stellar interiors and photon conversion in strong astrophysical magnetic fields. The MeV domain therefore represents the most promising observational window for testing this scenario. Future observations with improved sensitivity in the $(0.2-5)$ MeV range, together with improved modelling of neutron-star interiors, magnetospheric plasma distributions, magnetic-field geometries, and photon attenuation processes, will be essential to fully exploit this opportunity and to determine whether ALP-induced emission can be distinguished from conventional magnetospheric backgrounds.

Our study thus motivates dedicated searches for persistent MeV emission from young high-field pulsars and magnetar-like objects. A joint analysis of multiple sources and future MeV observations could provide complementary constraints on the ALP mass and photon coupling, potentially exploring regions of parameter space that remain inaccessible to current gamma-ray observations and laboratory searches.

%%%%%%%%%%%%%%%%%%%%%%%%%%%%%%%%%%%%%%%%%%

\section*{Acknowledgements}

We thank Sandro Mereghetti and David Smith for providing valuable insights into pulsar models. F. G., C. A. and MAPG acknowledge financial support by Junta de Castilla y León projects SA101P24, SA091P24, MICIU project PID2022-137887NB-I00, PID2025-168032NB-I00, Gravitational Wave Network (REDONGRA) Strategic Network (RED2024-153735-E) from Agencia Estatal de Investigación del MICIU (MICIU/AEI/10.13039/501100011033).

G.G. is supported by a contribution from Grant No. ASI-INAF 2023-17-HH.0.

BJK thanks the Spanish Agencia Estatal de Investigaci\'on (AEI, MICIU) for their support under the project \textsc{DMpheno2lab} (PID2022-139494NB-I00) financed by MCIN /AEI /10.13039/501100011033 / FEDER, EU; and the project \textsc{CosmoRadio} (PID2025-169666NB-I00) financed by MICIU/AEI/10.13039/501100011033 and by the FSE+.

\bibliographystyle{elsarticle-num} 
\bibliography{alpbibfile}

\end{document}